\documentclass[10pt,3p,review,authoryear]{elsarticle}

\usepackage[dvipsnames]{xcolor}
\usepackage{soul,framed} 
\usepackage{mathtools}
\usepackage{lscape}
\colorlet{shadecolor}{yellow}
\usepackage[ruled,vlined]{algorithm2e}
\usepackage{amsmath}
\usepackage{amsfonts}
\usepackage{amssymb}
\usepackage{subcaption}
\usepackage{hyperref}
\usepackage{bbm}
\usepackage{multirow}
\usepackage{booktabs}
\usepackage{pifont}

\usepackage{array}

\usepackage[utf8]{inputenc}
\usepackage{placeins}

\usepackage{blindtext}
\usepackage{dirtytalk}
\usepackage{adjustbox}
\usepackage{pdfpages} 
\usepackage{pgffor} 
\usepackage{xr}
\makeatletter
\AtBeginDocument{\let\LS@rot\@undefined}
\cleardoublepage\makeatother

\usepackage[shortlabels]{enumitem}
\usepackage{mdwtab}
\usepackage{eqparbox}
\usepackage{url}

\usepackage{amsthm}
\theoremstyle{plain}
\newtheorem{remark}{Remark}

\usepackage{mathtools}
\usepackage{dsfont}
\usepackage{bbold}
\usepackage[english]{babel} 
\usepackage{blindtext}
\usepackage{epstopdf}
\DeclareGraphicsExtensions{.png,.pdf,.jpg}
\usepackage{float}
\usepackage{hyperref}
\hypersetup{
    colorlinks=true,
    linkcolor=blue,
    filecolor=magenta,      
    urlcolor=cyan,
    pdftitle={Overleaf Example},
    pdfpagemode=FullScreen,
    }
\usepackage[utf8]{inputenc}
\usepackage{blindtext}

\newcommand{\PM}{\mathrm{PM}}
\newcommand{\NO}{\mathrm{NO}}

\begin{document}

\begin{frontmatter}

\title{E-STGCN: Extreme Spatiotemporal Graph Convolutional Networks for Air Quality Forecasting}

\author{Madhurima Panja\textsuperscript{1,}
\footnote[0]{\textit{Equal Contributions}}, 
Tanujit Chakraborty\textsuperscript{1, 0, 2}
\footnote[5]{\textit{Corresponding author}: \textit{Mail}: tanujit.chakraborty@sorbonne.ae},
Anubhab Biswas\textsuperscript{3},
Soudeep Deb\textsuperscript{4}\\
{\scriptsize \textsuperscript{1} Department of Science and Engineering, Sorbonne University Abu Dhabi, UAE.} \\
{\scriptsize \textsuperscript{2} Sorbonne Center for Artificial Intelligence, Sorbonne University, Paris, France.}\\
{\scriptsize \textsuperscript{3} University of Applied Sciences and Arts of Southern Switzerland, Switzerland.}\\
{\scriptsize \textsuperscript{4} Indian Institute of Management Bangalore, India.}}


\begin{abstract}
Modeling and forecasting air quality is crucial for effective air pollution management and protecting public health. Air quality data, characterized by nonlinearity, nonstationarity, and spatiotemporal correlations, often include extreme pollutant levels in severely polluted cities (e.g., Delhi, the capital of India). This is ignored by various geometric deep learning models, such as Spatiotemporal Graph Convolutional Networks (STGCN), which are otherwise effective for spatiotemporal forecasting. This study develops an extreme value theory (EVT) guided modified STGCN model (E-STGCN) for air pollution data to incorporate extreme behavior across pollutant concentrations. E-STGCN combines graph convolutional networks for spatial modeling and EVT-guided long short-term memory units for temporal sequence learning. Along with spatial and temporal components, it incorporates a generalized Pareto distribution to capture the extreme behavior of different air pollutants and embed this information into the learning process. The proposal is then applied to analyze air pollution data of 37 monitoring stations across Delhi, India. The forecasting performance for different test horizons is compared to benchmark forecasters (both temporal and spatiotemporal). It is found that E-STGCN has consistent performance across all seasons. The robustness of our results has also been evaluated empirically. Moreover, combined with conformal prediction, E-STGCN can produce probabilistic prediction intervals. 
\end{abstract}


\end{frontmatter}


\section{Introduction}

The rapid acceleration of industrialization and urbanization has spurred economic growth globally but has also exacerbated environmental issues, with air pollution ranking among the most pressing concerns \citep{brunekreef2002air, shaddick2020half}. According to the World Health Organization (WHO), approximately seven million premature deaths per year are linked to air pollution, a figure emphasizing the critical need for effective air quality management policies\footnote{\url{https://www.who.int/health-topics/air-pollution}}. The most harmful air pollutants include particulate matters ($\PM$), nitrogen dioxide ($\NO_2$), ozone ($\mathrm{O}_3$), sulfur dioxide ($\mathrm{SO}_2$), and carbon monoxide ($\mathrm{CO}$), which are closely monitored due to their impact on public health, particularly in terms of cardiovascular and respiratory diseases \citep{lelieveld2015contribution, olaniyan2020association}. Recognizing this threat, the United Nations has included air quality as a key component of its Sustainable Development Goals (SDGs)\footnote{\url{https://sdgs.un.org/goals}}. Likewise, the U.S. Environmental Protection Agency and equivalent authorities worldwide have implemented National Ambient Air Quality Standards (NAAQS) to limit pollution levels, which are crucial for safeguarding human health and the environment. Our focus in this article is on India, where the Central Pollution Control Board (CPCB) specifies that the hourly average concentrations of $\PM$ with a diameter of 2.5 micrometers or less ($\PM_{2.5}$) and $\PM$ with a diameter of 10 micrometers or less ($\PM_{10}$) pollutants should not exceed 60 micrograms per cubic meter ($\mu g/m^3$) and 100 $\mu g/m^3$, respectively\footnote{\url{https://cpcb.nic.in/upload/NAAQS_2019.pdf}}. However, in practice, air quality levels often surpass these predefined standards. Data from 37 monitoring stations (spatial locations) across Delhi, the capital city of India, collected over five years from 2019 to 2023\footnote{Delhi remains the world's most polluted capital city in 2023; \url{https://www.statista.com}.}, shows average $\PM_{2.5}$ and $\PM_{10}$ concentrations exceeding 100 $\mu g/m^3$ and 200 $\mu g/m^3$ respectively, significantly above the recommended limits. The situation deteriorates further with the onset of winter months due to low temperature and Delhi's landlocked geographical location, which hinders the dispersion of pollutants by wind. As a result, Delhi experiences a surge in pollution during winter, increasing the risk of chronic respiratory and cardiovascular diseases, neurological disorders, and a higher burden of mortality \citep{salvi2018burden}. \cite{pandey2021health} further highlights that air pollution adversely affects India’s economic growth as well.  Given these concerns, our study focuses on developing a spatiotemporal forecasting model to improve air quality predictions in urban environments. Such models are essential for informing public behavior and helping authorities implement timely interventions to mitigate health risks.

Research on air quality forecasting can broadly be classified into two categories: physical models and data-driven methods. Traditional physical models rely on fundamental principles of atmospheric science to simulate the emission, transport, and dispersion of pollutants within a target area. A couple of well-known methods in this category are the community multi-scale air quality \citep{byun2006review} and the nested air quality prediction model system \citep{wang2014modeling}, among many others. However, these often require extensive theoretical knowledge, carefully selected features, and region-specific parameters, prohibiting their usage in building a real-time air quality monitoring system. Other methods like Gaussian plume models or the Operational Street Canyon models lack the accuracy needed for real-time forecasting due to their reliance on limited parameters \citep{vardoulakis2003modelling, byun2006review}. In contrast, data-driven methods, which leverage historical information to capture pollution trends, have shown some promise \citep{lei2019macao}. Thus, traditional statistical models such as ARIMA (autoregressive integrated moving average) and dynamic factor models are widely used \citep[see, e.g.,][]{kumar2010arima}, but they are limited in their capacity to handle complex and nonlinear interactions in air quality data. To that end, recent advancements in machine learning, particularly deep learning, have significantly improved forecasting accuracy. For instance, \cite{li2017long} showed that long short-term memory (LSTM) networks can capture complex temporal dependencies and outperform traditional models in air quality forecasting. \cite{du2019deep} introduced a novel hybrid deep learning framework by combining bi-directional LSTM and one-dimensional convolutional neural networks (CNN). Extant literature on pollution forecasting also includes the use of recurrent neural networks \citep{ong2016dynamically}, transformers \citep{vaswani2017attention}, and temporal convolutional networks (TCN) \citep{samal2021multi}. However, a critical limitation of many deep learning models is their focus on temporal data alone, often overlooking spatial dependencies between monitoring locations. Since neighboring locations influence pollutant levels at any given station, a spatiotemporal approach is essential to model air pollution dispersion \citep{zhou2024predicting} accurately. Graph-based modeling strategies have gained significant attention in this regard, with graph neural networks (GNN) and graph convolutional networks (GCN) revolutionizing spatiotemporal forecasting \citep{scarselli2008graph}. In the current context, \cite{gao2021graph} leveraged GNNs with LSTMs to capture spatiotemporal information at the $\PM_{2.5}$ level. GCNs are also effective for air quality modeling, particularly due to their ability to perform convolutional operations that propagate information between nodes in a graph, thus leveraging localized aggregation of features from neighboring nodes \citep{Yu_2018}. For a comprehensive discussion in this context, refer to \cite{atluri2018spatio, jin2024survey}. 

The application of GCNs for modeling spatial dynamics of air pollution monitoring stations is, however, limited due to scalability and data sparsity issues. Moreover, existing forecasting architectures often struggle to accurately predict peaks in a time series, which is critical for air pollution forecasting to anticipate exceedances beyond regulatory thresholds and ensure ambient air quality standards are maintained. Due to the catastrophic nature of extreme values in air quality data, it is essential to understand and predict values that exceed the NAAQS threshold for developing effective early warning systems. For that, we turn attention to the extreme value theory (EVT) which provides a statistical framework for analyzing rare events, offering insights into the probability distribution of extreme pollutant concentrations \citep{coles2001introduction}. This framework has been successfully applied in various fields, including hydrology, climate studies, and air quality analysis, to predict the likelihood of exceeding established safety thresholds \citep{horowitz1980extreme, sharma1999application, ray2023pattern}. In air quality studies, EVT methods such as block maxima (BM) and peaks over threshold (POT) have been employed to model pollutant extremes  \citep{reiss1997statistical}. The probability distribution of these extreme events can be modeled using the generalized extreme value (GEV) distribution for BM and generalized Pareto (GP) distributions for the POT method. These techniques help in estimating the likelihood of extreme occurrences, allowing for the detection of potential rare events. EVT-based studies have been instrumental in forecasting pollution exceedances, thus supporting effective intervention strategies \citep{kan2004statistical, sfetsos2006extreme}. The reader is further referred to \cite{martins2017extreme} for a comprehensive review of EVT tools in air pollution problems. 

Interestingly, despite its proven utility, EVT has not been combined with spatiotemporal forecasting methods to build early warning systems for environmental preparedness. Our study aims to bridge this gap by introducing a novel EVT-guided modified spatiotemporal graph convolutional network (E-STGCN) model to handle the nonlinear, nonstationary, and extreme behavior for major air pollutants in Delhi, specifically $\PM_{2.5}$, $\PM_{10}$, and $\NO_2$. We examine the extreme behavior of these pollutants across 37 sensor locations using the POT method, modeled by the GP distribution. Integrating these insights into a modified version of spatiotemporal GCN (modified STGCN) enhances its ability to forecast air quality time series data while also capturing the extreme observations. This enables the E-STGCN model to capture extremes in comparison with standard spatiotemporal GCN (STGCN), which is designed for inference on graph structures with temporal dependencies in traffic flow applications \citep{Yu_2018}. Standard STGCN \citep{Yu_2018} leverages graph convolutions to model spatial dependencies and 1-D convolutional layers with gating mechanisms to capture short-term temporal patterns. On the contrary, the proposed E-STGCN framework builds upon the foundational spatial convolution blocks of STGCN; however, it differs from the temporal forecasting perspective and loss function construction. In the E-STGCN architecture, we retain the graph convolutional structure of STGCN to encode the spatial dependencies among monitoring stations effectively. However, we introduce two key innovations in the framework: (1) Employing LSTM units in place of temporal convolution layers to better capture long-memory temporal dynamics (evident in air pollutant time series), we name this architecture as modified STGCN and (2) Introducing a new hybrid loss function based on data loss and negative log-likelihood of the GP distribution (we call it POT loss). The use of the POT loss function enhances the modeling ability of rare but impactful extreme pollution events. The EVT component in E-STGCN enhances the model’s sensitivity to tail behavior via introducing a task-specific inductive bias that prioritizes accuracy in the extreme regime, which is typically underrepresented in conventional data loss formulations. There is a conceptual connection between the proposed E-STGCN and that of Physics-informed neural networks (PINN) \citep{raissi2019physics}, which combines (noisy) data with physical models and implements them through deep neural networks. However, in our method, instead of a physical model, we use an EVT-based GP distribution-fitted model as the building block for E-STGCN. Moreover, to evaluate the effect of the POT loss, we compare it with the modified version of STGCN and the standard STGCN model, proposed by \cite{Yu_2018}. Modified STGCN shares the same architecture as E-STGCN, including the graph convolutional and LSTM components, but is trained solely using the standard data loss, without incorporating the EVT module. As we shall demonstrate in our empirical study, the proposed E-STGCN, which enhances the original STGCN by incorporating both LSTM-based temporal modeling and an EVT-based loss function, achieves superior performance while capturing extreme air pollutant concentrations. In contrast, modified STGCN, which retains the architectural enhancements of E-STGCN but excludes the EVT component, shows improved baseline performance over STGCN in capturing long memory but lacks the same precision as E-STGCN in modeling extreme events. Furthermore, our proposed framework is scalable and capable of generating multi-step forecasts for both low and high-frequency spatiotemporal datasets. Unlike traditional GNN models, typically optimized for hourly predictions and shorter horizons (e.g., 12 to 72 hours), E-STGCN handles daily air quality data, providing reliable long-term forecasts at 30-, 60-, and 90-day horizons. Additionally, we apply conformal prediction methods to quantify forecast uncertainties, offering critical probabilistic insights for policy planning. 

To establish the efficacy of the proposed algorithm, we rigorously evaluate the model against state-of-the-art approaches. A list of these competing methods and their modeling capabilities is summarized in Table \ref{Tab_Baseline_Model_Capabilities}. Among time-dependent models, we consider the ubiquitous ARIMA approach \citep{box2015time} as well as several deep learning techniques, including LSTM \citep{hochreiter1997long}, TCN \citep{chen2020probabilistic}, DeepAR \citep{salinas2020deepar}, Transformers \citep{wu2020deep}, and NBeats \citep{oreshkin2019n}. For the spatiotemporal models, we evaluate the performance of Space-time Autoregressive Moving Average (STARMA) \citep{pfeifer1980three}, Generalized Space-time Autoregressive (GSTAR) \citep{cliff1975model}, Fast Gaussian Process (GpGp) \citep{guinness2018permutation}, STGCN \citep{Yu_2018}, Spatiotemporal Neural Network (STNN) \citep{saha2020hybrid}, modified STGCN (introduced here), and DeepKriging \citep{nag2023spatio}. In the interest of space, more details of these models are provided in Appendix A.2. 

\begin{table}[!ht]
\caption{Comparison of forecasting frameworks. The columns indicate whether each model can address spatiotemporal correlations, nonlinearity, and non-stationarity in time series data. Additional columns assess whether the method at its core can produce probabilistic forecasts, scalability (size of the data does not pose a problem), and handle extreme observations.}
  \centering {\tiny
  \begin{tabular}{lcccccc}
         \toprule
        Models &	Spatiotemporal &	Nonlinear &	Non-stationarity & Probabilistic & Scalability & Extreme Value\\ 
        &	&	&	& Forecasting & & Handling \\ \midrule
        ARIMA \citep{box2015time}
        & {\color{black}\ding{56}} 
        & {\color{black}\ding{56}} 
        & {\color{blue}\ding{52}} 
        & {\color{blue}\ding{52}} 
        & {\color{black}\ding{56}} 
        & {\color{black}\ding{56}} 
        \\
        LSTM \citep{hochreiter1997long}
        & {\color{black}\ding{56}} 
        & {\color{blue}\ding{52}} 
        & {\color{blue}\ding{52}} 
        & {\color{black}\ding{56}} 
        & {\color{blue}\ding{52}} 
        & {\color{black}\ding{56}} 
        \\
        TCN \citep{chen2020probabilistic}
        & {\color{black}\ding{56}} 
        & {\color{blue}\ding{52}} 
        & {\color{blue}\ding{52}} 
        & {\color{black}\ding{56}} 
        & {\color{blue}\ding{52}} 
        & {\color{black}\ding{56}} 
        \\
        DeepAR \citep{salinas2020deepar}
        & {\color{black}\ding{56}} 
        & {\color{blue}\ding{52}} 
        & {\color{blue}\ding{52}} 
        & {\color{black}\ding{56}} 
        & {\color{blue}\ding{52}} 
        & {\color{black}\ding{56}} 
        \\        
        Transformers \citep{wu2020deep}
        & {\color{black}\ding{56}} 
        & {\color{blue}\ding{52}} 
        & {\color{blue}\ding{52}} 
        & {\color{black}\ding{56}} 
        & {\color{blue}\ding{52}} 
        & {\color{black}\ding{56}} 
        \\
        NBeats \citep{oreshkin2019n}
        & {\color{black}\ding{56}} 
        & {\color{blue}\ding{52}} 
        & {\color{blue}\ding{52}} 
        & {\color{black}\ding{56}} 
        & {\color{blue}\ding{52}} 
        & {\color{black}\ding{56}} 
        \\
        STARMA \citep{pfeifer1980three}
        & {\color{blue}\ding{52}} 
        & {\color{black}\ding{56}} 
        & {\color{black}\ding{56}} 
        & {\color{blue}\ding{52}} 
        & {\color{black}\ding{56}} 
        & {\color{black}\ding{56}} 
        \\
        GSTAR \citep{cliff1975model}
        & {\color{blue}\ding{52}} 
        & {\color{black}\ding{56}} 
        & {\color{blue}\ding{52}} 
        & {\color{black}\ding{56}} 
        & {\color{black}\ding{56}} 
        & {\color{black}\ding{56}} 
        \\
        GpGp \citep{guinness2018permutation}
        & {\color{blue}\ding{52}} 
        & {\color{black}\ding{56}} 
        & {\color{black}\ding{56}} 
        & {\color{blue}\ding{52}} 
        & {\color{blue}\ding{52}} 
        & {\color{black}\ding{56}} 
        \\
        STGCN \citep{Yu_2018}
        & {\color{blue}\ding{52}} 
        & {\color{blue}\ding{52}} 
        & {\color{blue}\ding{52}} 
        & {\color{black}\ding{56}} 
        & {\color{blue}\ding{52}} 
        & {\color{black}\ding{56}} 
        \\
        STNN \citep{saha2020hybrid}
        & {\color{blue}\ding{52}} 
        & {\color{blue}\ding{52}} 
        & {\color{blue}\ding{52}} 
        & {\color{black}\ding{56}} 
        & {\color{black}\ding{56}} 
        & {\color{black}\ding{56}} 
        \\
        DeepKriging \citep{nag2023spatio}
        & {\color{blue}\ding{52}} 
        & {\color{blue}\ding{52}} 
        & {\color{blue}\ding{52}} 
        & {\color{blue}\ding{52}} 
        & {\color{black}\ding{56}} 
        & {\color{black}\ding{56}} 
        \\ \midrule
        Modified STGCN
        & {\color{blue}\ding{52}} 
        & {\color{blue}\ding{52}} 
        & {\color{blue}\ding{52}} 
        & {\color{blue}\ding{52}} 
        & {\color{blue}\ding{52}} 
        & {\color{black}\ding{56}} 
        \\
        Proposed E-STGCN 
        & {\color{blue}\ding{52}} 
        & {\color{blue}\ding{52}} 
        & {\color{blue}\ding{52}} 
        & {\color{blue}\ding{52}} 
        & {\color{blue}\ding{52}} 
        & {\color{blue}\ding{52}} 
        \\ \bottomrule
   \end{tabular}}
   \label{Tab_Baseline_Model_Capabilities}
\end{table}

The remainder of this paper is organized as follows. Section \ref{Sec_EVT} provides a brief description of extreme value theory. In Section \ref{Sec_Method}, we introduce the proposed E-STGCN architecture. Section \ref{Sec_Exp_Eval} outlines the experimental setup and reports the air quality forecasting results. In Section \ref{Sec_Discussion}, we discuss the implications of our approach to air quality forecasting. Finally, Section \ref{Sec_Conclu} concludes the paper and suggests future research directions.

\section{Preliminaries on Extreme Value Theory (EVT)}\label{Sec_EVT}

EVT focuses on analyzing the stochastic behavior of rare or extreme events within a given stochastic process. The goal of extreme value analysis is to quantify unusually large or small events and estimate the probability of these extreme occurrences, which differ significantly from the more common observations in the data. EVT deals with the asymptotic distribution of extreme order statistics, especially in the context of large datasets. This theory has been implemented in diverse domains, including earth sciences \citep{katz2002statistics}, economics and finance \citep{marimoutou2009extreme},  public health \citep{thomas2016applications}, and engineering \citep{castillo2012extreme}, among others. As mentioned above, statistical methods for modeling extreme events primarily rely on two approaches, i.e., block maxima and peaks over the threshold. Below, we briefly summarize the EVT methods that were utilized in this study. 

\subsection{Block Maxima (BM) Approach}

The BM method analyzes extreme events in a time series dataset \citep{gumbel1958statistics}. Given a sequence of time-dependent observations, this method divides the dataset into equal-sized non-overlapping blocks and considers the maximum value from each block as the extreme value of the time series. The probability distribution of the extremes is modeled using the generalized extreme value (GEV) distribution. To mathematically explain this, let $X_1, X_2, \ldots, X_n$ be independent and identically distributed (iid) random variables with continuous distribution function $F(\cdot)$. Then, as $n \longrightarrow \infty$, the distribution of $M_n = \max_{1 \leqslant i \leqslant n} X_i$ converges to $G\left(x\right)$, called the GEV distribution, defined by \citep[following][]{fisher1928limiting} 
\begin{equation*}
    G(x)=\begin{cases}
 \exp\left\{-\left(1+\xi_G\left(\frac{x-\mu_G}{\sigma_G}\right)\right)^{-1/\xi_G}\right\} & \text{if } \xi_G \neq 0, \\
 \exp\left\{-\exp\left(-\left(\frac{x-\mu_G}{\sigma_G}\right)\right)\right\} & \text{if } \xi_G=0.
\end{cases}
\end{equation*}
In the above distribution, $\xi_G \in\mathbb{R}$ is the extreme value index and it controls the shape of the distribution, $\mu_G \in\mathbb{R}$ is the location parameter, and $\sigma_G >0$ is the scale parameter. Depending on the tail behavior of the distribution, which is influenced by $\xi_G$, the GEV family can be classified into three extreme value distributions: Gumbel $(\xi_G = 0)$, Fréchet $(\xi_G > 0)$, and Weibull $(\xi_G < 0)$. While the Gumbel type distributions are suitable for modeling the extremes of the exponentially decaying-tailed distribution, Fréchet and Weibull families are the reference classes for the extremes of heavy-tailed and finite-tailed distributions, respectively \citep{rocco2014extreme}. 

Although the BM method has been widely used for extreme value analysis, it has several drawbacks. The partitioning of the dataset in this approach leads to significant information loss, as only the maximum value from each block is retained, potentially missing multiple extreme observations within a block. Also, usually, multiple extreme observations happen within a short time interval, which cannot be captured by the block maxima method. POT tries to overcome the disadvantages of the BM approach.

\subsection{Peaks Over Threshold (POT) Approach}\label{Sec_POT}

The POT approach is a key technique in EVT that identifies observations exceeding a pre-selected threshold, known as extreme values \citep{balkema1974residual}. By concentrating only on observations above the threshold, the POT approach offers an efficient and accurate mechanism for modeling tail behavior compared to conventional methods that assess the entire distribution. Given a time series dataset $\{e_1, e_2, \ldots, e_l\}$ and a threshold $\tau^*$ (any observations that exceed the threshold are called extreme events), the POT approach selects extreme events when $e_i > \tau^*$. The distribution of exceedances over the large threshold $\tau^*$ asymptotically follows a Generalized Pareto (GP) distribution. To explain it mathematically, let $\mathcal{Z}_1, \mathcal{Z}_2, \ldots, \mathcal{Z}_n$ be a series of iid random variables with a marginal distribution $Q(\cdot)$. \cite{pickands1975statistical} approximated the exceedance distribution for sufficiently large threshold values using a GP distribution, defined by 
\begin{equation}\label{POT_Eq}
    \mathcal{H}(z) = \begin{cases}
1-\left(1+\frac{\xi z}{\sigma}\right)^{-1/\xi} & \text{if } \xi\neq0\\
1-\exp\left(-\frac{z}{\sigma}\right) & \text{if } \xi=0.
\end{cases}
\end{equation}

Here, $\xi \in \mathbb{R}$ is the shape parameter and $\sigma > 0$ is the scale parameter of the GP distribution. The shape parameter $\xi$ plays a key role in determining the qualitative behavior of the GP distribution and influences its domain of attraction. When $\xi = 0$, $\mathcal{H}(z)$ belongs to the Gumbel distribution family, where the probability of extreme observations decreases exponentially, as indicated by its light tails. For $\xi > 0$, $\mathcal{H}(z)$ follows a Fréchet distribution characterized by heavy tails, suggesting more frequent extreme observations. Conversely, when $\xi < 0$, $\mathcal{H}(z)$ corresponds to a Weibull distribution with short tails, implying a lower probability of extreme observations. 

The POT approach offers a robust technique for effectively modeling extreme observations with minimum data loss. It is particularly suited for capturing the clustering effect, which is a prominent phenomenon in extreme events. The advantages of this method for modeling extreme air pollution levels are demonstrated in \cite{al2018modeling}, where the POT approach has been applied to investigate air pollution index exceedances in urban areas of Peninsular Malaysia.

\subsection{Methods for Threshold Selection in POT Approach}

The choice of threshold plays a key role in identifying the extreme observations in the dataset, thus significantly impacting the effectiveness of the POT approach. If a low threshold is selected, usual observations can be treated as extreme and violate asymptotic assumptions. On the contrary, a high threshold value can overlook potential extreme observations by treating too few data points as extreme. The threshold selection can be done objectively through a bias-variance trade-off or determined subjectively, with input from domain experts. Among various statistical procedures, the mean excess plot (MEP) is a popular approach for determining the threshold in the POT method \citep{benktander1960analytical}. The mean excess function of the random variable $\mathcal{Z}$ with distribution function $\mathcal{Q}_{\mathcal{Z}}(z)$ and right endpoint $z_R$ is given by
\begin{equation*}
    \operatorname{ME}\left(\tau^*\right) := E\left(\mathcal{Z} - \tau^* \mid \mathcal{Z} > \tau^*\right) = \int_{\tau^*}^{z_R} \left(\frac{1 - \mathcal{Q}_{\mathcal{Z}}(s)}{1 - \mathcal{Q}_{\mathcal{Z}}(\tau^*)}\right) \, ds,
\end{equation*}
provided $E\left(\mathcal{Z}\right) < \infty$ \citep{embrechts2013modelling}. Thus, if we model the statistical properties of exceedance for any arbitrarily chosen random variable $\mathcal{Z}$ among $\mathcal{Z}_1, \mathcal{Z}_2, \ldots, \mathcal{Z}_n$, with GP $(\sigma, \xi)$ distribution (as in Eq. \eqref{POT_Eq}), then the expected value of $\mathcal{Z}$ will be finite if and only if $\xi < 1$ and the mean excess function can be computed as:
\begin{equation*}
    \operatorname{ME}\left(\tau^*\right) = \frac{\sigma}{1 - \xi} + \frac{\xi}{1 - \xi}\tau^*,
\end{equation*}
where $0 \leqslant \tau^* < \infty$ if $0 \leqslant \xi < 1$ and $0 \leqslant \tau^* \leqslant -\frac{\sigma}{\xi}$ if $\xi < 0$. A natural estimate of the mean excess function, $\widehat{\operatorname{ME}}\left(\tau^*\right)$, is defined by
\begin{equation*}
    \widehat{\operatorname{ME}}\left(\tau^*\right) = \frac{\sum_{i = 1}^n \left(z_i - \tau^*\right) \operatorname{I}_{\left[z_i > \tau^*\right]}}{\sum_{i = 1}^n \operatorname{I}_{\left[z_i > \tau^*\right]}}; \; \tau^* \geqslant 0,
\end{equation*}
where $\operatorname{I}$ denotes the indicator function. This method considers the set of all points $\{(\tau^*, \widehat{\operatorname{ME}}(\tau^*)): \tau^* < z_{(n)}\}$ where $z_{\left(n\right)}$ is the highest order statistic from the sample. In principle, the MEP will appear linear if the exceedance observations are fitted with a GP distribution, which has a finite mean \citep{das2025pattern}. 

The MEP approach serves as an effective and objective tool for threshold selection in the absence of specific regulatory thresholds. For example, in a study on extreme influenza cases in Zhejiang, China, \cite{chen2015using} employed the MEP method to determine an optimal threshold for influenza incidence, owing to the lack of a standardized cutoff. This illustrates how MEP can successfully identify a threshold that balances the representation of both regular and extreme events. Similar applications can be found in fields such as hydrology \citep{durocher2019comparison}, finance \citep{chukwudum2020optimal}, and environmental science \citep{ghosh2010discussion}, where predefined regulatory thresholds are absent. In contrast, our study examines air pollution concentrations where public health and policy alignment are critical. The NAAQS, specified by the CPCB, offers scientifically established thresholds for air pollutant levels aimed at protecting public health. Although the MEP-derived threshold may be useful in the absence of regulatory cut-offs, it often corresponds to pollutant levels substantially higher than those defined by regulatory standards, reducing its practical utility. If the MEP threshold deviates significantly from the NAAQS, it may overlook numerous events of public health concern. Therefore, we adopt the NAAQS threshold to ensure that our findings align with public health policy and aid in designing intervention strategies.

\section{Proposed Methodology}\label{Sec_Method}
This section introduces the proposed E-STGCN along with the modified STGCN method for spatiotemporal forecasting of air pollution concentration levels in the presence of extreme observations. Specifically, Section \ref{Sec_Problem_Formulation} outlines the mathematical formulation of the spatiotemporal air pollution forecasting problem, while Section \ref{Sec_Model_Overview} provides an overview of the E-STGCN architecture, with detailed descriptions of the components within each module of the proposed framework.

\subsection{Problem Formulation} \label{Sec_Problem_Formulation}

In this study, we address the air quality prediction problem as a spatiotemporal forecasting task, where the key challenge is to model temporal patterns from historical data while simultaneously capturing the spatial relationships between multiple air quality monitoring stations. Let there be $N$ monitoring stations and suppose $\mathbf{X}_t = \left[X_t^1, X_t^2, \ldots, X_t^N \right] \in \mathbb{R}^N$ denote the vector of air pollutant concentrations across all stations at time $t$, where $t = [1, T]$. Our objective is to generate $q$-step-ahead ($q \geqslant 1$) forecasts $\left\{\widehat{\mathbf{X}}_{T+1}, \widehat{\mathbf{X}}_{T+2}, \ldots, \widehat{\mathbf{X}}_{T+q}\right\}$ based on $T$ past observations $\mathcal{X}_{1:T} = \left[\mathbf{X}_1, \mathbf{X}_2, \ldots, \mathbf{X}_T\right]^\text{T} \in \mathbb{R}^{T \times N}$ which represents the air pollutant concentration across $T$ historical observations and $N$ monitoring stations. To achieve this, we develop a forecasting model that integrates extreme value theory with GCN and LSTM networks for modeling the spatiotemporal correlations among the $N$ monitoring stations, accounting for the presence of extreme observations in the dataset. Our goal is to learn a forecasting function $F_{\operatorname{E-STGCN}}$ that maps spatiotemporal inputs into a sequence of future predictions. To understand the objective formally, let $G = \left\{V, E\right\}$ denote an undirected graph, where $V \in \mathbb{R}^N$ represents the set of nodes, corresponding to the monitoring stations and the set of edges $E \in \mathbb{R}^{N \times N}$ indicates the spatial correlations between the stations. Mathematically, the air quality forecasting problem can then be expressed as:
\begin{equation*}
    F_{\operatorname{E-STGCN}} : \left(\mathcal{X}_{1:T}, A\right) \rightarrow \left\{\widehat{\mathbf{X}}_{T+1}, \widehat{\mathbf{X}}_{T+2}, \ldots, \widehat{\mathbf{X}}_{T+q} \right\}
\end{equation*}
where the adjacency matrix $A \in \mathbb{R}^{N \times N}$ encode the spatial connections and   $\widehat{\mathbf{X}}_{T+i} = \{\widehat{X}^1_{T+i}, \widehat{X}^2_{T+i}, \ldots, \widehat{X}^N_{T+i}\} \in \mathbb{R}^N$ represents the $i$-step-ahead ($i = [1, q]$) forecast of air pollution concentrations for the $N$ monitoring stations, computed based on the $T$ historical observations.

\subsection{E-STGCN Model Overview} \label{Sec_Model_Overview}

\begin{figure}[!ht]
    \centering
    \includegraphics[width = \textwidth]{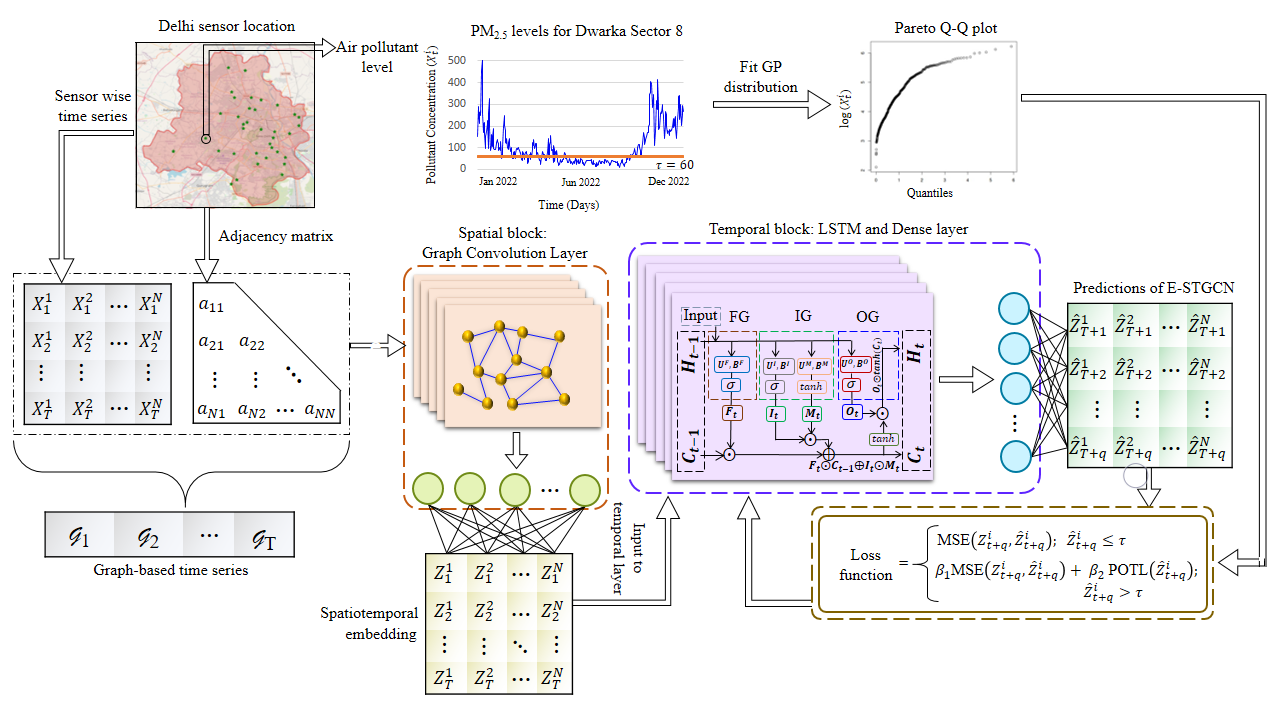}
    \caption{\textbf{Extreme Spatiotemporal Graph Convolutional Networks (E-STGCN)}. Daily air pollution concentration levels from different regions of Delhi, along with the corresponding adjacency matrix, are processed through a Graph Convolutional Network (GCN) and a dense layer to generate spatiotemporal embeddings. To account for extreme values, each sensor's time series data is modeled using a Generalized Pareto (GP) distribution. The GCN-embedded data is then passed through an LSTM layer, followed by a dense layer, to produce accurate forecasts. The network is trained using a modified loss function that combines the conventional mean squared error (MSE) loss with a peaks-over-threshold loss (POTL) when predictions exceed a predefined threshold ($\tau$).}
    \label{figModel_Architecture}
\end{figure}

The overall architecture of the E-STGCN framework, depicted in Fig.\ \ref{figModel_Architecture}, consists of three primary modules: the spatial module, the temporal module, and the EVT module. The spatial module maps the input data onto attributed spatiotemporal graphs and learns the underlying spatial correlations. These learned graph structures and historical air pollutant concentrations are processed through the spatial blocks comprising GCNs and fully connected neural networks, which capture dynamic temporal information and spatial influences. The output from the spatial module is then fed into the temporal module, where the future trajectories of air pollutant concentrations are predicted using recurrent LSTM layers and a fully connected dense layer (modified to capture long-memory dynamics and differs from standard STGCN). The EVT module, another key component of the E-STGCN architecture, is specifically designed to model rare but impactful extreme pollution events. It fits a GP distribution to the historical air pollutant concentrations that exceed permissible thresholds. This fitted distribution is used to augment the conventional data-driven loss with POT loss (discussed in Section \ref{Sec_EVT_Module}). The hybrid loss function ensures that predictions exceeding the regulatory thresholds are not only penalized for their squared error but are also encouraged to conform to the statistical structure of the learned tail distribution. The primary objective of E-STGCN is to capture extremes, not merely to be robust to them. Specifically, the model aims to accurately forecast both the occurrence and magnitude of rare pollution events, a crucial drawback of prior spatiotemporal deep learning models. The EVT module directly addresses this gap by embedding domain-specific tail behavior into the training process. By integrating EVT-based statistical knowledge with the spatiotemporal representations learned from the spatial and temporal modules, E-STGCN can effectively model the dynamics of air pollutant concentrations, particularly in scenarios involving threshold exceedances. Our framework is conceptually aligned with the idea of PINN \citep{raissi2019physics, karniadakis2021physics}, where domain knowledge is embedded into neural networks via loss constraints (not architectural changes). PINN uses physical laws to inform model behavior in regions where data is scarce or uncertain by taking the total loss as:
$
\operatorname{Loss}_{\operatorname{total}} =  \operatorname{Loss}_{\operatorname{data}} + \tilde{\lambda} \operatorname{Loss}_{\operatorname{physics}},
$
where $\operatorname{Loss}_{\operatorname{data}}$ is the data-centric loss and $\operatorname{Loss}_{\operatorname{physics}}$ is the residual from the governing ordinary differential equations or partial differential equations. In E-STGCN, we use EVT to regularize the model where the data is sparse in the tail (certainly many extreme values). In this way, the GP distribution penalty serves as a probabilistic prior grounded in statistical theory, analogous to physical laws in the mechanistic model, thus allowing E-STGCN to accurately forecast extreme observations in pollution concentration levels.

\subsubsection{Spatial Module}

In the spatial domain, the air pollutant concentrations at different sensor locations influence each other with varying intensities, and most interactions are dynamic. To capture the spatial correlations among the monitoring stations, we employ graph convolution operations. Typically, GCNs allow convolution operations on arbitrary graph structures, enabling the learning of node-order invariant representations. In the E-STGCN framework, we model the historical air pollutant concentrations using GCN by considering the geographical locations of the monitoring stations as nodes, which form the basis of spatial dependencies. Thus, the undirected graph $G = \left\{V, E\right\}$, with $V$ nodes and $E$ connecting edges, can be represented using an adjacency matrix $A$ for efficient computer processing. Specifically, the adjacency matrix $A$ is static and is constructed based on the weighted Haversine distance ($d_{ij}$) between the geographical locations of the $i^{th}$ station (with latitude $\phi^{i}$, longitude $\lambda^{i}$) and the $j^{th}$ station as:
\begin{equation} \label{Eq_Haversine_distance}
d_{ij}=2R \sin^{-1}\left[\sqrt{\sin^2\left(\frac{\Delta\phi}{2}\right) + \cos\left(\lambda^i\right) \cos\left(\lambda^j\right)\sin^2\left(\frac{\Delta\lambda}{2}\right)}\right],
\end{equation}
where $\Delta\phi = \phi^i-\phi^j$, $\Delta\lambda = \lambda^i-\lambda^j$, and $R$ represents the earth's radius. The weighted adjacency matrix, indicating the similarity between the corresponding nodes, are computed using a Gaussian kernel as 
\begin{equation}\label{Adjacency_Eq}
a_{ij}=\exp\left(-\frac{d^2_{ij}}{\tilde{\sigma}^2}\right), \text{ when } \; i\neq j \text{ and } \exp\left(-\frac{d^2_{ij}}{\tilde{\sigma}^2}\right)\geqslant \epsilon,
\end{equation}
where $\tilde{\sigma}^2$ and $\epsilon$ are the parameters that control the distribution and sparsity of the adjacency matrix $A$. Specifically, if the distance between the nodes exceeds $\sqrt{-\tilde{\sigma}^2 \ln \epsilon}$, no edges are considered between the nodes. 

With a slight abuse of terminology, let us represent the air pollutant concentrations monitored at $N$ stations over $T$ timestamps as spatiotemporal graphs $\mathcal{G} = \left \{\mathcal{G}_1, \mathcal{G}_2, \ldots, \mathcal{G}_T \right\}$, where each graph $\mathcal{G}_t = \left\{\mathbf{X}_t, A\right\}$ consists of $\mathbf{X}_t \in \mathbb{R}^N$, representing the pollutant attributes at time $t$, and $A \in \mathbb{R}^{N \times N}$, providing the structural information for the $N$ stations. To map the non-Euclidean spatiotemporal graphs to spatiotemporal node embeddings, we perform localized convolutions of the node neighborhood using GCN layers. GCNs generalize the concept of CNN filters to graph-structured data by applying polynomial filters over neighboring nodes. These filters can be approximated using Chebyshev polynomials of order $d$ as:
\begin{equation*}
    \mathcal{P}_w(L) = \sum_{u = 0}^{d} w_u \mathcal{C}_u (\tilde{L}),
\end{equation*}
where $\mathcal{C}_u$ denotes the $u^{t h}$ Chebyshev polynomial and $\tilde{L} = \frac{2L}{\zeta_{max}} - I_N$ is the normalized graph Laplacian. Here, $L=D-A$ is the Laplacian, $D$ is the diagonal degree matrix with $D_{ii} = \sum_j A_{ij}$, and $\zeta_{max}$ is the largest eigenvalue of $L$. The graph convolution on input data $\mathbf{X}_t$ is then given by:
\begin{equation}\label{Eq_graph_Conv}
\mathbf{X}_t^{'} = \mathcal{P}_w(L) \mathbf{X}_t. 
\end{equation}	
As shown by \cite{kipf2016semi}, a first-order approximation ($d=1$) is often sufficient and computationally efficient, leading to a simplified graph convolution. Hence, Eq. \eqref{Eq_graph_Conv} can be simplified as:
\begin{equation}
    \mathbf{X}_t^{'} = w_0 \mathbf{X}_t + w_1 \left(\frac{2L}{\zeta_{max}} - I_N\right) \mathbf{X}_t,
\end{equation}
where $w_0$ and $w_1$ are filter weights shared across all $N$ nodes. By applying a stack of $K$ different polynomial filter layers, which corresponds to a sequence of $K$ graph convolution layers with $1^{st}$-order approximations, the spatiotemporal node embeddings for $\mathcal{G}_t = \left\{\mathbf{X}_t, A\right\};$ $t = 1,2, \ldots, T$, can be computed as:
\begin{align*}
    & h_t^{i, (0)} = X_t^i \\
    & h_t^{i, (k)} = f_t^{(k)} \left( W_t^{(k)}  \frac{\sum_{j \in \mathcal{N}(i)} h_t^{j, (k-1)}}{|\mathcal{N}(i)|}  + B_t^{(k)} h_t^{i, (k-1)}\right); k = 1, 2, \ldots K\\
   & Z_t^i = \operatorname{Dense}\left(h_t^{i, (K)}\right),
\end{align*}
where in the $k^{th}$ iteration, the function $f_t^{(k)}$ and filter weights $\{W_t^{(k)}, B_t^{(k)}\}$ are shared to update the initial embedding using 1-hop localized convolutions, repeated $K$ times based on a neural message passing mechanism \citep{gilmer2017neural}. Thus, $h_t^{i, (k)}$ is the embedding of node $i$ at timestamp $t$ during iteration $k$, computed by taking the mean of its neighboring nodes and its self-embedding from the previous iteration at time $t$. The final spatiotemporal representation, $Z_t^i$, from the spatial block is computed by modeling the GCN output from the $K^{th}$ layer using a fully connected dense layer. Consequently, the spatiotemporal embedding $\mathbf{Z}_t = \left[Z_t^1, Z_t^2, \ldots, Z_t^N\right] \in \mathbb{R}^N$ generated from $\mathcal{G}_t$ updates $\mathbf{X}_t$ with the encoded information from $(K-1)$-order neighborhood of the central node through $K$ successive filtering operations. It is of the essence here to point out that the $1^{st}$-order approximation of the polynomial filter is highly effective and scalable for large-scale graph structures \citep{Yu_2018}.

\subsubsection{Temporal Module}

The temporal module of the E-STGCN framework is designed to model the spatiotemporal embeddings, $\left[\mathbf{Z}_1, \mathbf{Z}_2, \ldots, \mathbf{Z}_T\right]^\text{T}$, learned in the spatial block. Due to the complex sequential dependencies within $\left\{\mathbf{Z}_t\right\}$, we employ an LSTM network, a robust variant of RNNs, that efficiently overcomes optimization challenges of conventional RNN architectures \citep{hochreiter1997long}. This temporal module differs from the architecture of the standard STGCN, where 1-D convolutional layers with gating mechanisms are used. The LSTM network in E-STGCN introduces specialized memory cells that replace standard hidden units, improving stability, speed, and accuracy. These cells maintain self-connected recurrent edges with fixed weights, enabling stable gradient flow across long sequences. The cell state stores long-term information, while hidden states manage short-term context. A gating mechanism, comprising forget, input, and output gates, controls updates to both states, allowing the network to effectively capture and retain long and short-term dependencies in sequential data. At each timestamp $t$, for each of the $N$ nodes, the LSTM receives $p$ lagged values $\underbar z_t^i = \{Z_{t-p+1}^i, Z_{t-p+2}^i,\ldots, Z_{t}^i\} \in \mathbb{R}^p$ along with the previous hidden state vector $H_{t-1}^i$ as input. It then generates the $q$-steps-ahead projections of $Z_{t}^i$ along with the new memory $M_t^i$, updating both the hidden state $H_t^i$ and the cell state $C_t^i$. The forget gate determines which parts of the previous cell state $C_{t-1}^i$ should be retained by computing the forget gate activation as:
\begin{equation*}
    F_t^i = \phi_1\left(U^i_{ZF} \underbar z_t^i + U^i_{HF} H_{t-1}^i + B^i_F\right),
\end{equation*}
where $U^i_{ZF} \in \mathbb{R}^{m \times p}, U^i_{HF} \in \mathbb{R}^{m \times m}, \text{ and } B^i_{F} \in \mathbb{R}^{m}$ are learnable parameters, $m$ is the number of hidden layers, and $\phi_1$ is a sigmoid function that constrains values of $F_t^i \in [0, 1]$. A value of $F_t^i$ close to 1 retains the corresponding part of $C_{t-1}^i$; while a value close to 0 discards it. The input gate controls how much new information from the current input $\underbar z_t^i$ should influence the cell state by computing the activation ($I_t^i$ ) and new memory vector $(M_t^i)$ as:
\begin{equation*}
    I_t^i = \phi_1\left(U^i_{ZI} \underbar z_t^i + U^i_{HI} H_{t-1}^i + B^i_I\right), \quad
    M_t^i = \phi_2\left(U^i_{ZM} \underbar z_t^i + U^i_{HM} H_{t-1}^i + B^i_M\right),
\end{equation*}
where $\phi_2$ is tangent hyperbolic activation function, $U^i_{ZI}, U^i_{ZM}, U^i_{HI}, U^i_{HM}, B^i_{I}, \text{ and } B^i_{M}$ are learnable parameters. Using $F_t^i, I_t^i, \text{ and } M_t^i$ current cell state is updated as:
\begin{equation*}
   C_{t}^i = F_t^i \odot C_{t-1}^i \oplus I_t^i \odot M_t^i,
\end{equation*}
where $\odot$ denotes the element-wise multiplication. Finally, the current hidden state is calculated in the output gate based on the activation vector $\left(O_{t}^i \right)$ of the output gate as:
\begin{equation*}
    H_t^i = O_t^i \odot \phi_2 \left(C_t^i\right), \quad \text{where }
    O_t^i = \phi_1\left(U^i_{ZO} \underbar z_t^i + U^i_{HO} H_{t-1}^i + B^i_O\right),
\end{equation*}
with $U^i_{ZO}, U^i_{HO}, \text{ and } B^i_{O}$ 
being the output gate parameters. To compute $H_t^i$ and $C_t^i$, the initial values are set to $H_0^i = C_0^i = 0$. Consequently, the $q$-steps-ahead forecast of the air pollutant concentrations for the $i^{th}$ node is obtained using a fully connected dense layer as:
\begin{equation*}
    \left\{\widehat{Z}_{t+1}^i, \widehat{Z}_{t+2}^i, \ldots, \widehat{Z}_{t+q}^i \right\} = \operatorname{Dense}\left(H_t^i\right).
\end{equation*}

The final output generated by the temporal module effectively captures the sequential patterns of the air pollutant series. However, the model struggles to forecast sudden peaks, which are particularly common in Delhi's air pollutant concentrations during winter months. To address this, we design the EVT module within the E-STGCN architecture, enabling the framework to accurately forecast spatiotemporal dependencies in situations of threshold exceedances.

\subsubsection{EVT Module}\label{Sec_EVT_Module}

In the field of air pollution control, \cite{roberts1979review} emphasized that rare events often hold more significance than regular observations. Therefore, prior knowledge of these rare occurrences is crucial for accurate modeling and forecasting of air pollutant concentrations. The spatial and temporal modules of the E-STGCN architecture leverage historical pollutant data from various proximal monitoring stations to predict future trends. However, their inability to differentiate between common and rare events limits their effectiveness in modeling extreme occurrences. To address this issue, the EVT module, a key component of the E-STGCN framework, utilizes extreme value theory to identify the underlying patterns of air pollutant concentrations associated with rare observations.

In the EVT module, we employ the POT approach to analyze the extreme observations and integrate them into the spatiotemporal forecasts of the previous modules. In the POT method (as discussed in Section \ref{Sec_POT}), we examine the behavior of exceedances by fitting a GP distribution to the pollutant concentrations that exceed the NAAQS threshold ($\tau$). Following Eq. \eqref{POT_Eq}, the conditional GP distribution for the $i^{th}$ monitoring station at time $t$ can be mathematically formulated as:
\begin{equation*}
    P\left[X_t^i - \tau \leqslant x_t^i \mid X_t^i \geqslant \tau \right] =  \begin{cases}
       1 - \left(1 + \frac{\xi^i x_t^i}{\sigma^i}\right)^{-1/\xi^i} & \text{if } \xi^i \neq 0 \\
       1 - \exp \left(-\frac{x_t^i}{\sigma^i}\right) & \text{if } \xi^i = 0,
    \end{cases} 
\end{equation*}
where $\xi^i \in \mathbb{R}$ is the shape parameter and $\sigma^i > 0$ is the scale parameter for the GP distribution fitted to the pollutant concentrations of the $i^{th}$ monitoring station. The shape parameter $\xi^i$ is particularly important as it influences the tail behavior of the GP distribution. To estimate the shape and scale parameters of the GP distribution for the $i^{th}$ station, we consider the $\tilde{k}^i$ observations $\left\{X_1^i, X_2^i, \ldots, X_{\tilde{k}^i}^i\right\}$ exceeding the threshold $\tau$ and compute the log-likelihood function based on the threshold exceedance as follows:
$$
l\left(\sigma^i, \xi^i\right) = - \tilde{k}^i \log\left(\sigma^i\right) - \left(1 + \frac{1}{{\xi}^i}\right) \sum_{m = 1}^{ \tilde{k}^i} \log\left(1 + \frac{{\xi}^ix_m^i}{{\sigma}^i}\right),
$$
provided $\left(1 + \frac{{\xi}^ix_m^i}{{\sigma}^i}\right) > 0 \text{ for } m = 1, 2, \ldots, \tilde{k}^i$; otherwise $l\left(\sigma^i, \xi^i\right) = - \infty$ \citep{grimshaw1993computing, coles2001introduction}. Since the log-likelihood of the GP distribution lacks a closed-form analytical solution, we employ the Broyden–Fletcher–Goldfarb–Shanno (BFGS) algorithm, a quasi-Newton optimization method, to estimate the parameters $\sigma^i$ and $\xi^i$ \citep{fletcher2000practical}. The BFGS algorithm iteratively updates parameter estimates using gradient information and approximates the inverse Hessian to maximize the log-likelihood, while ensuring stability and convergence by enforcing constraints such as the positivity of $\sigma^i$ and validity of the likelihood domain. The resulting estimates $\left(\hat{\sigma}^i, \hat{\xi}^i\right)$ of the scale $\left({\sigma}^i\right)$  and shape $\left({\xi}^i\right)$ parameters obtained from the air pollutant concentration levels of the $i^{th}$ monitoring station enables the design of the POT loss function which serves as a prior information in the E-STGCN framework. 

The POT loss function is computed as the negative log-likelihood of the fitted GP distribution. Now using a single predicted value of air pollution concentration level ($\widehat{Z}_{t}^i$) for monitoring station $i$ at time $t$ that exceeds the threshold $\tau$, we compute the negative log-likelihood (NLL) of the fitted GP distribution (POT loss) as:
\begin{equation}\label{Eq_POTL}
    \operatorname{POTL}\left(\widehat{Z}_{t}^i\right) = \operatorname{NLL} = \log\left(\hat{\sigma}^i\right) + \left(1 + \frac{1}{\hat{\xi}^i}\right) \log\left(1 + \frac{\hat{\xi}^i\widehat{Z}_{t}^i}{\hat{\sigma}^i}\right).
\end{equation}

We then incorporate the negative log-likelihood function (Eq. \eqref{Eq_POTL}) while designing the loss function of the temporal module. This approach enhances the modeling of threshold exceedances by incorporating knowledge from EVT as prior information to the model.

\subsubsection{Optimization}

The objective function of the E-STGCN framework is formulated as a combination of the data loss, computed by the mean squared error (MSE), and the POT-based loss function (Eq. \eqref{Eq_POTL}) depending on whether predicted values exceed a specified threshold $\tau$. This allows the E-STGCN to capture the dynamics of air pollution, not just the average behavior, but especially its rare and high-impact extremes (ones that violate regulatory thresholds by NAAQs). The hybrid loss function, combining MSE for general predictive accuracy with negative log-likelihood penalty from a GP distribution for threshold exceedance, in the form of POT loss, can be expressed as:
\begin{equation} \label{Temporal_loss_Eq}
    \operatorname{Loss}\left(Z_{t+q}^i, \widehat{Z}_{t+q}^i \right) = \begin{cases}
    \operatorname{MSE}\left(Z_{t+q}^i, \widehat{Z}_{t+q}^i \right), & \widehat{Z}_{t+q}^i \leqslant \tau \\
    \beta_1 \operatorname{MSE}\left(Z_{t+q}^i, \widehat{Z}_{t+q}^i \right) + \beta_2 \operatorname{POTL}\left(\widehat{Z}_{t+q}^i\right), & \widehat{Z}_{t+q}^i > \tau,        
    \end{cases}
\end{equation}
where $\beta_1$ and $\beta_2$ are the hyperparameters that regulate the contributions of data loss and the POT-based loss, respectively. Since the loss function is differentiable almost everywhere, we utilize the backpropagation method to train the corresponding weights. Thus, for predictions below the threshold $\tau$, the model is optimized solely based on MSE loss that draws the predictions toward the conditional mean. This is appropriate for the bulk of the data. However, for predictions above the threshold $\tau$, the loss function becomes: $\beta_1 \operatorname{MSE}\left(Z_{t+q}^i, \widehat{Z}_{t+q}^i \right) + \beta_2 \operatorname{POTL} \left(\widehat{Z}_{t+q}^i\right)$, where the model is encouraged not just to minimize MSE but to conform to the statistical structure of extreme values learned from the data. Note that the GP distribution penalty term does not force the prediction to match the mean of the GP distribution, but it acts like a regularization term, encouraging the prediction to fall within the plausible tail shape defined by EVT. This acts as a distribution-aware regularization, but not a replacement for the predictive loss. Using only the GP distribution likelihood for exceedance would ignore the distance between observed and predicted values; therefore, it can significantly harm the point prediction accuracy (which is crucial in air pollution forecasting for policy intervention). Hence, the combined loss provides a trade-off where the model learns to be accurate (stay close to the observed values) while regularizing the tail behavior via alignment with the empirical tail shape of the pollutant distribution. Additionally, to ensure an effective balance between data-based learning and the EVT knowledge, the coefficients $\beta_1$ and $\beta_2$ in the modified loss function are selected through a cross-validation approach by minimizing the root mean square error on the hold-out validation set (as indicated in Fig.\ \ref{Train_Test_Split}). These hyperparameters govern the trade-off between accurately forecasting average concentration levels and effectively capturing the extreme behavior of the dataset. As a result, the E-STGCN model minimizes a modified loss function that captures the underlying spatiotemporal dynamics while also integrating distributional characteristics through the POT-based component to better model extreme concentration levels. A detailed visualization showcasing the working principle of the E-STGCN framework is provided in Fig.\ \ref{figModel_Architecture}. Although designed for air pollution, the proposed hybrid loss function is model-agnostic and can be integrated into other spatiotemporal architectures for extreme forecasting tasks (e.g., finance, epidemics, etc.).

\begin{remark}
E-STGCN extends the foundational principles of the STGCN framework \citep{Yu_2018} by adopting a spatiotemporal learning paradigm that combines graph-based spatial modeling with temporal sequence processing. Both architectures utilize graph convolutional layers to capture spatial dependencies among monitoring stations using a predefined adjacency structure and are designed to forecast multivariate signals over a graph topology. Despite this shared foundation, E-STGCN introduces several important modifications, particularly in its temporal modeling strategy and training objective (loss function), that distinguish it from the original STGCN. The key differences are outlined below:
\begin{enumerate}
    \item \textit{Temporal Module:} STGCN uses 1-D convolutional layers coupled with Gated Linear Units, which are effective at capturing local temporal patterns within a fixed window. However, this design is limited in its ability to model long-term memory dependencies, which are frequently observed in air quality time series due to seasonal variation, meteorological shifts, and policy interventions. To address these limitations, E-STGCN replaces the temporal convolutional block with LSTM units (sequence-to-sequence architecture), which are specifically designed to capture (both short and long-memory) dynamic temporal dependencies and preserve contextual information over extended horizons. This modification significantly improves classical STGCN’s ability to model persistent pollution episodes and non-stationary temporal dynamics.
    \item \textit{Tail Modeling and Novel Loss Function:} E-STGCN incorporates a POT approach to explicitly separate moderate and extreme events, based on NAAQs regulatory thresholds. This allows the model to learn distinct dynamics for each regime and better capture extreme pollution events, which often exhibit different statistical behavior from regular observations. 
    E-STGCN introduces a hybrid loss function that combines the traditional MSE loss with a negative log-likelihood penalty derived from the GP distribution (POT loss), forming an EVT-based loss component. This additional term is activated only when the predicted values exceed a specified threshold $\tau$, guiding the model to produce forecasts that are not only accurate on average but also statistically consistent with the empirical tail behavior of the data. 
    In contrast, standard STGCN \citep{Yu_2018} lacks any mechanism to explicitly handle extreme events and treats all prediction errors uniformly, regardless of their magnitude or severity. This EVT-guided learning mechanism of E-STGCN enables it to generate risk-aware forecasts, a critical gap in STGCN's design.
\end{enumerate}

\end{remark}

\section{Experimental Evaluation}\label{Sec_Exp_Eval}

\begin{figure}[!ht]
    \centering
    \includegraphics[width = \textwidth]{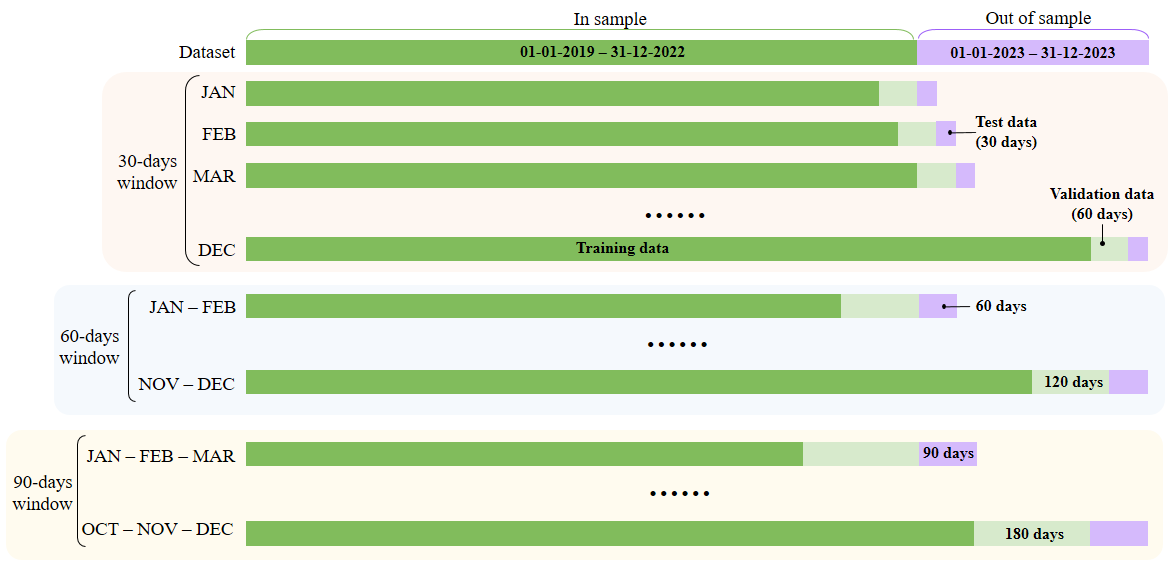}
    \caption{Dataset split for different forecast evaluation windows}
    \label{Train_Test_Split}
\end{figure}

In this study, we assess the efficiency of the proposed E-STGCN framework by comparing its forecasting performance with several temporal and spatiotemporal forecasters. We use daily data on Delhi's air pollutant concentration levels from January 1, 2019, to December 31, 2022, to train the models and generate forecasts for different months of 2023. To demonstrate the generalizability of our proposal, we evaluate its forecasting performance across three forecast horizons, namely short-term, medium-term, and long-term, spanning over 30 days, 60 days, and 90 days, respectively, using a rolling window approach. For the short-term horizon, forecasts are computed for each of the 12 months of 2023, separately. The forecast window covers two consecutive months in the medium-term horizon, resulting in 6 cases. There are four forecast windows for the long-term horizon, each covering three successive months. Fig.\ \ref{Train_Test_Split} visually represents the training, validation, and test periods used in the forecasting tasks. The following subsections present a brief description of the air pollutant datasets and their global characteristics (Section \ref{Section_Data_characterestics}), extreme value analysis of air pollutant concentrations (Section \ref{Sec_EVT_Modeling}), performance comparison metrics (Section \ref{Section_Performance_Metrics}), implementation of the proposed framework and experimental results (Section \ref{Section_Experimental_results}), statistical significance tests of the experimental results (Section \ref{Sec_statistical_significance}), and uncertainty quantification of the proposal (Section \ref{Sec_uncertainty_quantification}).

\subsection{Data and Preliminary Analysis}\label{Section_Data_characterestics}


In this study, we focus on forecasting the daily concentration levels of three major air pollutants, namely $\PM_{2.5}$, $\PM_{10}$, and $\NO_2$, and analyze their statistical and global features using the data collected from 37 monitoring stations located in Delhi. Pollution concentrations fluctuate significantly throughout the year across various stations, with ranges between $0.08 - 761.95 \, \mu g/m^3$ for $\PM_{2.5}$, $1.00 - 923.70 \, \mu g/m^3$ for $\PM_{10}$, and $0.13 - 428.15 \, \mu g/m^3$ for $\NO_2$. The average concentrations are 102.31 $\mu g/m^3$, 203.48 $\mu g/m^3$, and 43.02 $\mu g/m^3$ for $\PM_{2.5}$, $\PM_{10}$, and $\NO_2$ respectively. We also compute the five-point summary statistics, standard deviation (sd), coefficient of variation (cv), skewness, and kurtosis for the pollutant concentrations monitored at different stations. Furthermore, we analyze various global time series features, including long-term dependency, stationarity, linearity, and seasonality for the pollutant levels. The results of the descriptive statistics and global features of $\PM_{2.5}$, $\PM_{10}$, and $\NO_2$ datasets, as reported in Tables A.1, A.2, A.3 of the appendices, reveal that the air pollutant series from most of the monitoring stations exhibit long-range dependencies, non-stationary behavior, and nonlinear patterns. Additionally, some datasets display weekly and quarterly seasonality. 

\begin{figure}[!ht]
    \centering
    \includegraphics[width=\textwidth]{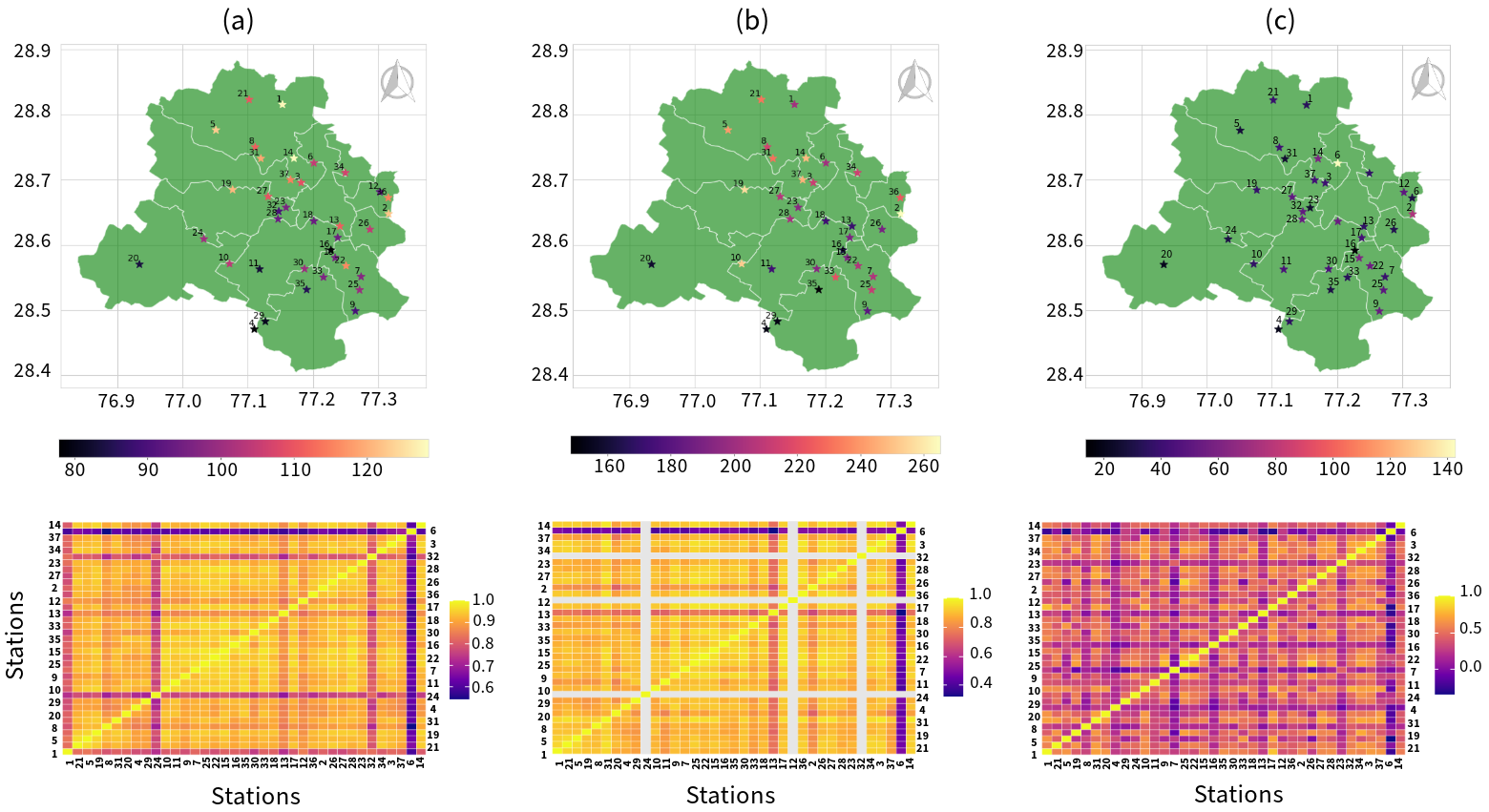}
    \caption{Upper panel: Spatial distribution of the monitoring stations in Delhi and average pollution level of (a) PM$_{2.5}$, (b) PM$_{10}$, and (c) NO$_{2}$. The lower panel represents the pairwise correlation between pollutant levels from each station for (a) PM${2.5}$, (b) PM${10}$, and (c) NO${2}$. Stations are reordered such that geographically proximate locations appear adjacent to each other, emphasizing correlations driven by spatial proximity. For PM$_{10}$, stations 12, 24, and 32 did not record data and are represented as white cells in the heatmap to denote missing observations.}
    \label{fig:Corr_Plot}
\end{figure}

Next, in Fig.\ \ref{fig:Corr_Plot}, we visualize the spatial distribution of the air pollutant monitoring stations in Delhi. The upper panel of the plot showcases the average pollution levels observed at each station. These plots highlight that monitoring stations in close proximity tend to record similar pollution concentration levels compared to distant ones, underscoring the spatial dependencies in pollutant concentrations. The lower panel of Fig.\ \ref{fig:Corr_Plot} displays the pairwise correlation heatmap of pollutant concentrations across monitoring stations. In the plot, the stations are clustered based on geographical proximity, using hierarchical clustering on their Haversine distance, as in Eq. \eqref{Eq_Haversine_distance}. This reordered matrix reveals spatial patterns of the pollution dynamics in the local neighborhood. The resulting heatmap demonstrates clear spatial dependency such that geographically closer stations tend to exhibit stronger correlations, owing to localized pollutant dispersion and common emission sources. The diagonal entries of the plot, representing self-correlations, are uniformly equal to 1. Notably, $PM_{2.5}$ and $PM_{10}$ concentrations show consistently strong positive correlations across most station pairs, while $NO_{2}$ exhibits moderate to strong correlations, primarily among nearby stations. Moreover, some geographically distant station pairs (e.g., stations 9 and 18; stations 21 and 23) also display high correlations, suggesting the influence of non-local emission sources. These spatial correlation structures reveal the complex interplay of local and regional factors influencing air quality in Delhi and provide critical insights for modeling both spatial and temporal dependencies in the forecasting process.

\subsection{Extreme Value Modeling of Air Quality Data}\label{Sec_EVT_Modeling}

\begin{figure}[!ht]
    \centering
    \includegraphics[width=\textwidth]{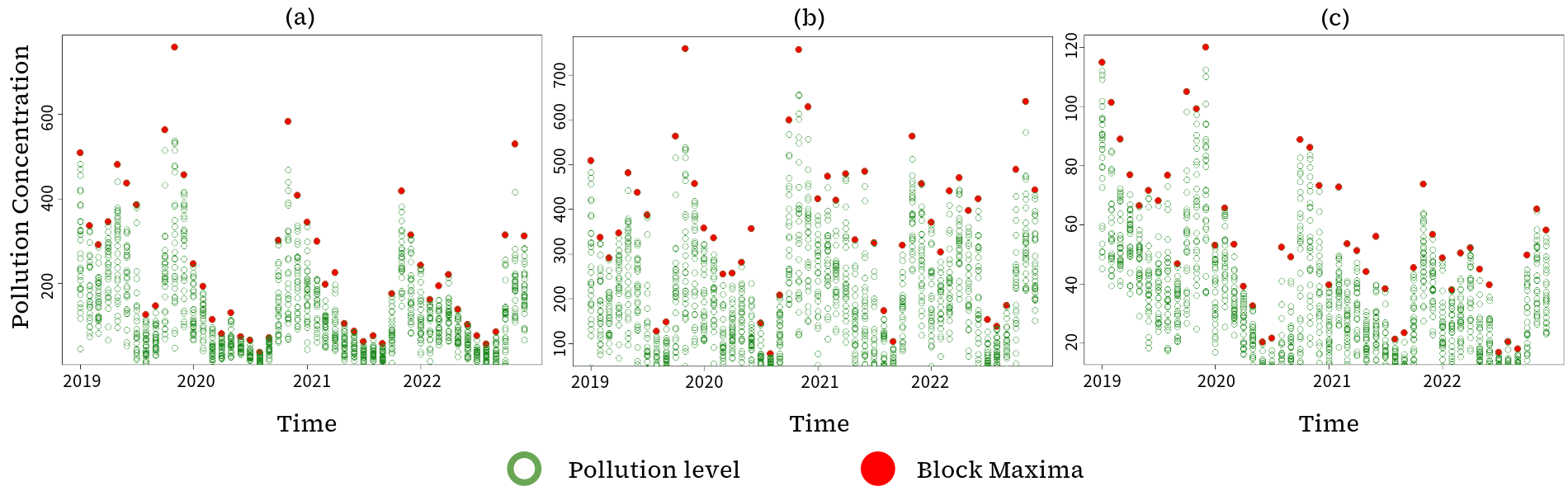}
    \caption{Block maxima plot for extreme value analysis of (a) $\PM_{2.5}$, (b) $\PM_{10}$, and (c) $\NO_{2}$ pollutant concentration in Alipur, Delhi monitoring station with each month representing a block. Green points indicate the pollution levels, and red circles are the maximum values identified for each block.}
    \label{fig:Block_Maxima_Plot}
\end{figure}

\begin{figure}[!ht]
    \centering
    \includegraphics[width=0.9\textwidth]{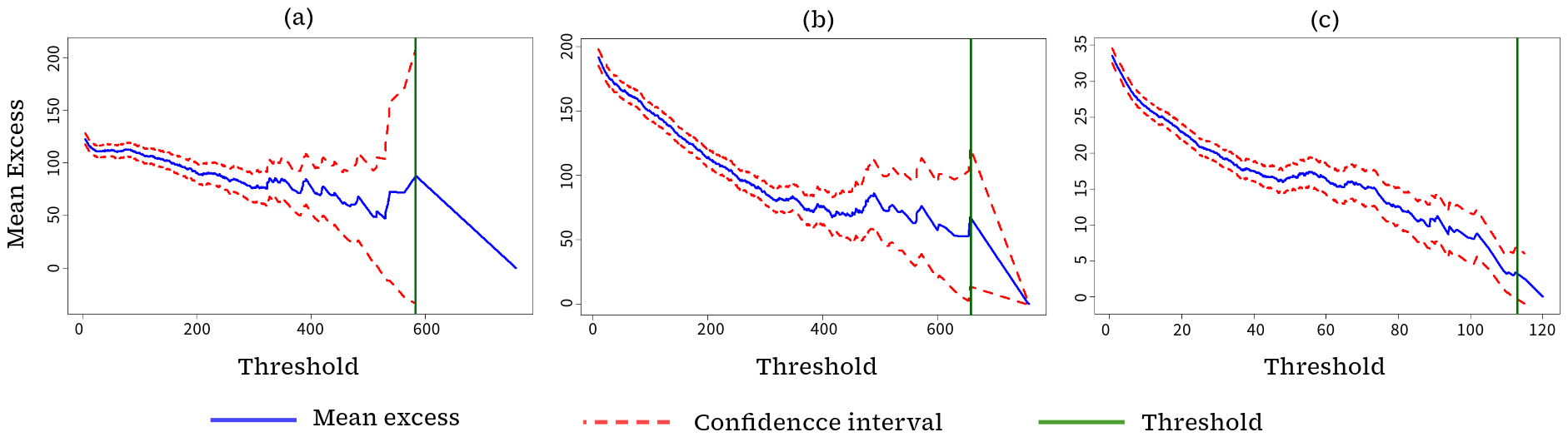}
    \caption{Mean excess plot for (a) $\PM_{2.5}$, (b) $\PM_{10}$, and (c) $\NO_{2}$ pollutant concentration in Alipur, Delhi monitoring station. The blue solid line indicates the mean excess level, the red dotted lines represent the 95\% confidence interval, and the green solid line is the threshold obtained from the mean excess plot.}
    \label{fig_Mean_Excess_Plot}
\end{figure}

In this section, we employ the BM method and the POT approach to detect and model the extreme observations in the air pollutant concentrations. Fig.\ \ref{fig:Block_Maxima_Plot} presents the results of extreme value analysis using the BM method for daily pollution concentrations of $\PM_{2.5}$, $\PM_{10}$, and $\NO_{2}$, measured at the Alipur monitoring station from 2019 to 2022. The other stations display similar behavior as well. In our analysis, we consider a block size of 30 days, representing the maximum value in each block with a red circle and the remaining observations with green circles. From the plots, it can be observed that in several blocks, the maximum values are not necessarily extreme. Conversely, in other blocks, multiple extreme values, apart from the maximum, are abandoned by this method. To address these limitations, we employ the POT approach in our study. For determining the optimal threshold in the POT method, we utilize the MEP approach and demonstrate the results for pollution concentrations of $\PM_{2.5}$, $\PM_{10}$, and $\NO_2$ monitored at the same station in Fig. \ref{fig_Mean_Excess_Plot}. The plot highlights the mean excess value for various thresholds ($\tau^*$) with a 95\% confidence interval. From the MEP, we can observe that the mean excess value becomes linear beyond the green straight line, indicating that the corresponding value represents the threshold. Specifically, the MEP-based thresholds are 583 for $\PM_{2.5}$, 658 for $\PM_{10}$, and 116 for $\NO_2$ datasets. However, using these thresholds results in only 0.14\% extreme values for $\PM_{2.5}$, $\PM_{10}$, and $\NO_2$ dataset, which is insufficient for effective POT analysis. Nevertheless, in scenarios where extremely high observations are scarce, the MEP technique for threshold selection within the POT framework can be useful. For example, in case of $NO_2$ concentration levels from selected low-pollution monitoring stations in Delhi, specifically Station 4 (Aya Nagar), Station 16 (Lodhi Road IMD), and Station 20 (Najafgarh) of Delhi, where the frequency of extreme $NO_2$ levels ranges between 0.07\% – 0.21\% (as reported in Table A.3 of Appendix), the MEP approach can be utilized. Fig. \ref{MEP_NO2_Revised} displays the mean excess value computed for these datasets across varying thresholds ($\tau^*$), along with 95\% confidence intervals. The estimated MEP-based thresholds are 72 $\mu g/m^3$ for Station 4, 74 $\mu g/m^3$ for Station 16, and 80.5 $\mu g/m^3$ for Station 20, all of which closely align with the NAAQs regulatory threshold of 80 $\mu g/m^3$ for daily $NO_2$ exposure. These results demonstrate the practical validity of the MEP approach in relatively clean environments, where exceedance events are rare but still important to model. They further highlight the effectiveness of MEP in identifying reliable thresholds for tail modeling, even in low-exceedance regimes where capturing extreme behavior remains critical and a regulatory threshold is absent (e.g., epidemic datasets).

\begin{figure}[!ht]
    \centering
    \includegraphics[width=0.9\textwidth]{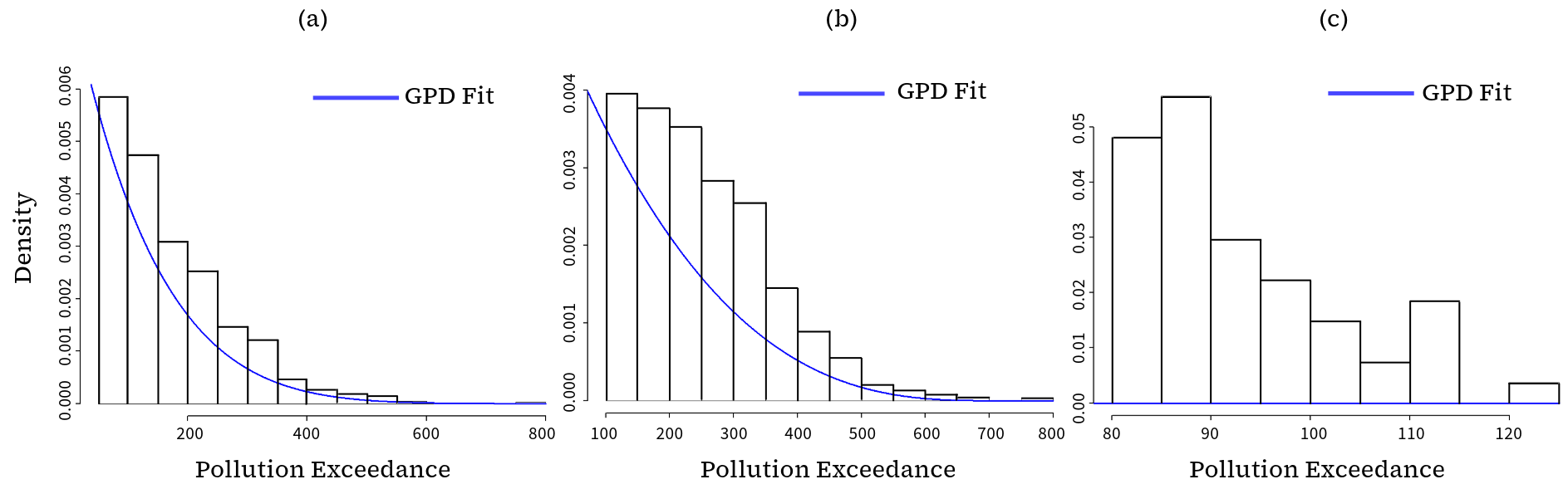}
    \caption{(a)-(c) Probability density plots of $\PM_{2.5}$, $\PM_{10}$, and $\NO_{2}$ pollutant concentration extremes in Alipur, Delhi monitoring station, respectively. All histograms are fitted with the probability density (blue) of the generalized Pareto distribution.}
    \label{fig:GPD_Fit_Extreme}
\end{figure}

In this study, we opt for a subjective method of threshold selection, utilizing the NAAQS established by the CPCB for industrial, residential, rural, and other areas.  Domain experts determine these thresholds to protect public health, vegetation, and the environment. Following the NAAQS recommendation, we set the threshold values as 60 $\mu g/m^3$ for $\PM_{2.5}$, 100 $\mu g/m^3$ for $\PM_{10}$, and 80 $\mu g/m^3$ for $\NO_{2}$ pollutants and examine the exceedance of pollution concentration levels over these thresholds. From Tables A.1, A.2, A.3 of Appendix A.1, the average exceedance levels are 61\% for $\PM_{2.5}$, 77\% for $\PM_{10}$, and 10\% for $\NO_2$.  Additionally, to verify the iid assumption of the POT approach for these exceedance datasets, we perform the Durbin-Watson (DW) test \citep{durbin1971testing}, which detects autocorrelation at lag 1 in the residuals from the regression analysis. The DW test p-values (refer to the above-mentioned tables in Appendix A.1) indicate that for most exceedance time series, lag 1 residuals are uncorrelated. However, for certain stations with limited observations above the threshold, the DW test statistic could not be computed. We also demonstrate the fitting of the GP distribution for different air pollutant concentrations with the selected thresholds in Fig.\ \ref{fig:GPD_Fit_Extreme}. Due to the absence of enough extreme observations in the $NO_2$ dataset, the GP distribution does not provide a good fit as opposed to the $PM_{2.5}$ and $PM_{10}$ datasets. 

\begin{figure}
    \centering
    \includegraphics[width=\linewidth]{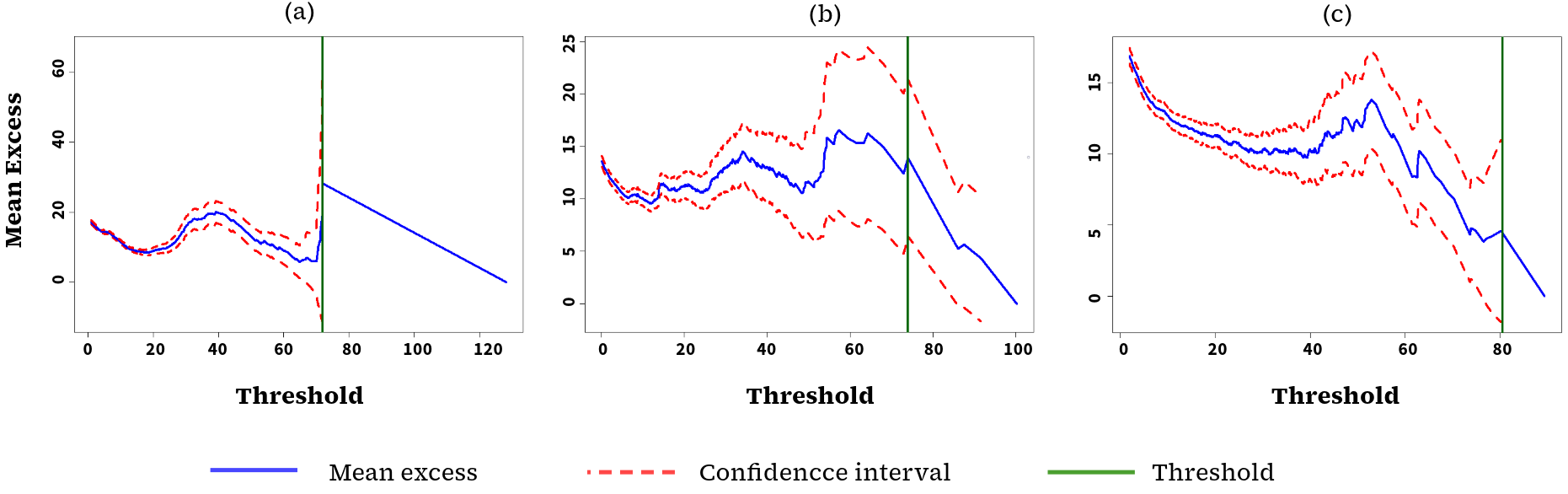}
    \caption{Mean excess plot for $\NO_{2}$ pollutant concentration monitored in (a) Station 4 (Aya Nagar), (b) Station 16 (Lodhi Road IMD), and (c) Station 20 (Najafgarh) of Delhi. The blue solid line indicates the mean excess level, the red dotted lines represent the 95\% confidence interval, and the green solid line is the threshold obtained from the mean excess plot.}
    \label{MEP_NO2_Revised}
\end{figure}

\subsection{Forecasting Performance Evaluation Measures} \label{Section_Performance_Metrics}

In our experimental evaluation, we utilize four deterministic and two probabilistic forecast performance indicators. The deterministic measures, including Mean Absolute Error (MAE), Mean Absolute Scaled Error (MASE), Root Mean Squared Error (RMSE), and Symmetric Mean Absolute Percent Error (SMAPE), quantify the average performance of different forecasters under various error formulations \citep{hyndman2018forecasting}. Additionally, to evaluate how well E-STGCN captures extreme pollution behavior, we include two probabilistic metrics, such as Continuous Ranked Probability Score (CRPS) and Pinball loss (quantile loss).  These metrics provide insight into the model’s ability to forecast not only central tendencies but also distributional tail characteristics, aligning with the goals of strictly proper scoring rules as emphasized by \cite{gneiting2007strictly}. Specifically, the quantile-based Pinball Loss evaluates performance at different quantile levels, while the CRPS assesses the overall quality of the predicted cumulative distribution. The mathematical formulations of these metrics are as follows:

{
\[\text{ MAE} = \frac{1}{q}\sum_{t =1}^{q} |X^i_t - \widehat{X}^i_t|, \; \;
\text{ MASE} = \frac{\sum_{t = T + 1}^{T+q} |\widehat{X}^i_t - X^i_t|}{\frac{q}{T-1} \sum_{t = 2}^T |X^i_t - X^i_{t-1}|}, 
\;\]
\[ \text{ RMSE} = \sqrt{\frac{1}{q}\sum_{t =1}^{q} (X^i_t - \widehat{X}^i_t)^2}, \; \;
\text{ SMAPE} = \frac{1}{q} \sum_{t=1}^q \frac{2|\widehat{X}^i_t - X^i_t|}{|\widehat{X}^i_t|+ |X^i_t|} \times 100 \%,
\]
\[ \text{ Pinball Loss}(\rho^*) = \frac{1}{q}\sum_{t =1}^{q} \operatorname{max}\left(\rho^*\left(X^i_t - \widehat{X}^i_{t}\right), (1-\rho^*)\left(X^i_t - \widehat{X}^i_{t}\right)\right), \; \text{and} \;
\]
\[\text{ CRPS} = \frac{1}{q} \sum_{t=1}^q \left[\int_{-\infty}^{\infty} \left(F_t^i\left(\tilde{u}\right) - \mathbb{1}\left(\tilde{u} \geqslant X^i_t\right) \right)^2d\tilde{u}\right],\]
}
where $q$ denotes the forecast horizon, $\rho^*$ is the quantile, $\widehat{X}^i_t$ is the forecast of the actual value $X^i_t$, $F_t^i$ is the predicted probability distribution function (PDF) for the $i^{th}$ station at time $t$, and $T$ is the size of the training sample. In our analysis, we set $\rho^* = 0.8$ and report the values of the probabilistic metrics based on the 80\% quantile \citep{gneiting2023model}. By definition, the minimum value of these performance measures suggests the `best-fitted' model. 

\subsection{Experimental Setup and Forecasting Accuracy}\label{Section_Experimental_results}

In this section, we discuss the implementation of the proposed E-STGCN approach for forecasting air pollutant concentrations in Delhi. To train the sequential workflow of our model, we first utilize the `fgpd' function from the \textit{evmix} package in R. This function computes the maximum likelihood estimates for the scale parameter ($\sigma^i>0$) and the shape parameter ($\xi^i \in \mathbb{R}$) of the GP distribution, based on the training dataset for the $i^{th}$ station whenever an exceedance over the NAAQS threshold occurs. These estimated parameters provide prior information regarding extreme values in the training data. Subsequently, we implemented the E-STGCN model in Python to generate the spatiotemporal forecasts for the proposed approach. For modeling the spatial dependencies in the dataset, we compute the adjacency matrix ($A$) based on the weighted Haversine distance, as in Eq. \eqref{Adjacency_Eq}. This matrix identifies the neighbors for each sensor, organizing their locations into a graphical structure by identifying relevant nodes and edges. Next, we employ CNNs and a dense layer from the \textit{TensorFlow} library to encode the training data's structural and feature-based information. To model the temporal dependencies, the output of the spatial module is passed through an LSTM layer and a dense layer. The weights of the temporal layer are optimized using a custom loss function, which combines the MSE loss with a POT-based loss (as in Eq. \eqref{Temporal_loss_Eq}). This modified loss function leverages prior information about the NAAQS exceedances to enhance the accuracy of air pollution forecasts. Once the E-STGCN model and other benchmark forecasters are implemented, we generate out-of-sample forecasts using a rolling window approach for different forecast horizons. Below, we summarize the performance of our proposal and baseline models from temporal and spatiotemporal paradigms based on several key performance indicators. A brief description of the benchmark temporal and spatiotemporal baseline models used in the experimental analysis, along with their implementation details, is outlined in Appendix A.2. 

Tables \ref{Tab_PM25_30_Eval}, \ref{Tab_PM10_30_Eval}, and \ref{Tab_NO2_30_Eval} present the performance of the proposed model and the baseline architectures in generating short-term forecasts for $\PM_{2.5}$, $\PM_{10}$, and $\NO_2$ levels, respectively. As indicated in Table \ref{Tab_PM25_30_Eval}, the proposed E-STGCN model achieves state-of-the-art performance for several months of 2023. In particular, during the onset (November) and end (February) of winter, our proposed framework generates the most accurate forecasts and efficiently captures the overall variability, including extreme observations, for the $\PM_{2.5}$ concentration levels. While NBeats and ARIMA outperformed in December and January, the E-STGCN model improved ARIMA’s forecast by 22.4\% (based on the MAE metric) in March, and its performance remained competitive with NBeats. In April, our proposal shows a similar trend by efficiently capturing the extreme values in the dataset, while its average performance remains comparable to ARIMA and NBeats. During the summer months, from May to September, the proposed framework outperformed the benchmark models across both deterministic and probabilistic performance indicators, except in July, when LSTM produced more accurate average forecasts and E-STGCN provided the most precise predictions for extreme values. In October, the GpGp and the Transformers framework recorded the lowest forecast error. For short-term forecasting of $\PM_{10}$, the E-STGCN model consistently performed best for the first three months of 2023, as measured by most point-based and distribution-based performance metrics. During April and May, the $\PM_{10}$ concentration levels rarely exceeded the NAAQs threshold, leading to similar performance between the E-STGCN and modified STGCN models, as the framework was trained primarily on the data loss. The average performance of the STNN model is better during June; however, the E-STGCN framework provides a better performance in capturing the tail distribution of the data. In July and December, although the NBeats framework outperforms all the forecasting approaches, our proposed model regains its forecasting superiority from August to October, as indicated by the average and distributional error metrics. In November, the spatiotemporal GpGp model showed competitive performance with our method. For 30-day ahead forecasts of $NO_2$ concentration levels, the E-STGCN and modified STGCN models showed similar performance. Since the average exceedance of $NO_2$ levels over the NAAQS threshold was around 10\%, the use of a POT-based loss function was limited. Nevertheless, both architectures outperformed the standard STGCN, underscoring the effectiveness of incorporating an LSTM-based temporal module in capturing long-memory dynamics of air pollutant datasets. As shown in Table \ref{Tab_NO2_30_Eval}, the E-STGCN and modified STGCN models provided the lowest forecast errors in several months, including January, February, April, May, and August to November. For March and December, the STARMA, ARIMA, and NBeats frameworks outperform the competitive forecasting models, whereas in July, the ESTCGN framework accurately predicts the tail behavior of the dataset.

\begin{table}[!ht]
    \centering
    \footnotesize
    \caption{Forecasting performance of the proposed E-STGCN model in comparison to the temporal-only and spatiotemporal forecasting techniques for 30 30-day ahead forecast horizon of PM$_{2.5}$ pollutant (best results are \underline{\textbf{highlighted}}).}
    \resizebox{\textwidth}{!}{
\begin{tabular}{|c|c|cccccc|ccccccc|c|}
\hline Forecast & \multirow{3}{*}{Metric} & \multicolumn{6}{c|}{Temporal-only Model}                  & \multicolumn{7}{c|}{Spatiotemporal Model}           & Proposed \\ \cline{3-15}
     \multirow{2}{*}{Period} & & \multirow{2}{*}{ARIMA} &  \multirow{2}{*}{LSTM} & \multirow{2}{*}{TCN} & \multirow{2}{*}{DeepAR} & \multirow{2}{*}{Transformers} & \multirow{2}{*}{NBeats} & \multirow{2}{*}{STARMA} & \multirow{2}{*}{GSTAR} & \multirow{2}{*}{GpGp} & \multirow{2}{*}{{\color{black}STGCN}} & \multirow{2}{*}{STNN} & Modified & \multirow{2}{*}{DeepKrigging} & \multirow{2}{*}{E-STGCN} \\ 
    & & &  &  &  &  &  &  &  &  &  &  &  STGCN &  & \\ \hline                                 
\multirow{6}{*}{JAN} & MAE    & \underline{\textbf{54.88}}               & 160.54 & 108.88 & 164.82 & 93.72        & 61.72  & 95.87                 & 63.48  & 86.09 & {\color{black}66.51} & 65.90  & 69.11  & 179.16 & 56.07 \\
                     & MASE   & \underline{\textbf{0.96}}                & 2.85   & 1.94   & 2.93   & 1.63         & 1.10   & 1.67                  & 1.10   & 1.49 & {\color{black}1.16}  & 1.18   & 1.20   & 3.21   & 0.98             \\
                     & RMSE   & \underline{\textbf{71.18}}               & 174.93 & 137.98 & 178.87 & 114.86       & 78.46  & 112.01                & 82.27  & 107.61 & {\color{black}84.94} & 84.55  & 90.21  & 192.41 & 75.34            \\
                     & SMAPE  & \underline{\textbf{31.20}}               & 152.10 & 79.80  & 160.70 & 60.40        & 36.50  & 69.40 & 39.50  & 53.30 & {\color{black}39.01}  & 37.50  & 41.50  & 187.10 & 31.60            \\
\hline      
\end{tabular}
}
\label{Tab_PM25_30_Eval}
\end{table}

\begin{table}[!ht]
    \centering
    \footnotesize
    \caption{Forecasting performance of the proposed E-STGCN model in comparison to the temporal-only and spatiotemporal forecasting techniques for 30 30-day ahead forecast horizon of PM$_{10}$ pollutant (best results are \underline{\textbf{highlighted}}).}
    \resizebox{\textwidth}{!}{
\begin{tabular}{|c|c|cccccc|ccccccc|c|}
\hline Forecast & \multirow{3}{*}{Metric} & \multicolumn{6}{c|}{Temporal-only Model}                  & \multicolumn{7}{c|}{Spatiotemporal Model}           & Proposed \\ \cline{3-15}
    \multirow{2}{*}{Period} & & \multirow{2}{*}{ARIMA} &  \multirow{2}{*}{LSTM} & \multirow{2}{*}{TCN} & \multirow{2}{*}{DeepAR} & \multirow{2}{*}{Transformers} & \multirow{2}{*}{NBeats} & \multirow{2}{*}{STARMA} & \multirow{2}{*}{GSTAR} & \multirow{2}{*}{GpGp} & \multirow{2}{*}{{\color{black}STGCN}} & \multirow{2}{*}{STNN} & Modified & \multirow{2}{*}{DeepKrigging} & \multirow{2}{*}{E-STGCN} \\ 
    & & &  &  &  &  &  &  &  &  &  &  &  STGCN &  & \\ \hline
    
\multirow{6}{*}{JAN}             & MAE                     & 99.31  & 276.07 & 262.61 & 275.41 & 165.16       & 87.24  & 121.40 & 82.99  & 86.54 & {\color{black} 94.94}  & 115.31 & 89.89  & 233.97 & \underline{\textbf{82.67}}                             \\
                                 & MASE                    & 1.23   & 3.46   & 3.31   & 3.45   & 2.05         & 1.10   & 1.49   & \underline{\textbf{1.02}}   & 1.09 & {\color{black} 1.18}  & 1.47   & 1.12   & 2.94   & 1.03                              \\
                                 & RMSE                    & 126.85 & 294.14 & 307.81 & 293.53 & 192.87       & 112.07 & 147.81 & 110.33 & 109.86 & {\color{black} 123.03} & 145.59 & 119.16 & 273.87 & \underline{\textbf{108.41}}                            \\
                                 & SMAPE                   & 35.90  & 175.50 & 124.70 & 174.60 & 70.50        & 30.80  & 47.10  & 29.70  & 30.30 & {\color{black} 34.22}  & 37.00  & 32.00  & 120.60 & \underline{\textbf{29.10}}              \\
\hline                          
\end{tabular}
}
\label{Tab_PM10_30_Eval}
\end{table}

\begin{table}[!ht]
    \centering
    \footnotesize
    \caption{Forecasting performance of the proposed E-STGCN model in comparison to the temporal-only and spatiotemporal forecasting techniques for 30 30-day ahead forecast horizon of NO$_{2}$ pollutant (best results are \underline{\textbf{highlighted}}).}
    \resizebox{\textwidth}{!}{
\begin{tabular}{|c|c|cccccc|ccccccc|c|}
\hline Forecast & \multirow{3}{*}{Metric} & \multicolumn{6}{c|}{Temporal-only Model}                  & \multicolumn{7}{c|}{Spatiotemporal Model}           & Proposed \\ \cline{3-15}
    \multirow{2}{*}{Period} & & \multirow{2}{*}{ARIMA} &  \multirow{2}{*}{LSTM} & \multirow{2}{*}{TCN} & \multirow{2}{*}{DeepAR} & \multirow{2}{*}{Transformers} & \multirow{2}{*}{NBeats} & \multirow{2}{*}{STARMA} & \multirow{2}{*}{GSTAR} & \multirow{2}{*}{GpGp} & \multirow{2}{*}{{\color{black}STGCN}} & \multirow{2}{*}{STNN} & Modified & \multirow{2}{*}{DeepKrigging} & \multirow{2}{*}{E-STGCN} \\ 
    & & &  &  &  &  &  &  &  &  &  &  &  STGCN &  & \\ \hline
                                 
\multirow{6}{*}{JAN} & MAE    & 13.76               & 23.48 & 38.72  & 26.73  & 16.49        & 15.71  & 17.87                 & 15.76 & 16.37 & {\color{black} 17.68} & 28.46  & \underline{\textbf{12.28}} & 40.02  & \underline{\textbf{12.28}}            \\
                     & MASE   & 3.33                & 8.21  & 15.68  & 9.85   & 11.36        & 4.58   & 4.99                  & \underline{\textbf{2.48}}  & 4.48 & {\color{black} 10.59} & 8.15   & 4.74  & 13.43  & 4.74             \\
                     & RMSE   & 17.01               & 26.50 & 41.37  & 29.63  & 19.45        & 19.34  & 21.80                 & 19.37 & 19.92 & {\color{black}20.64} & 36.38  & \underline{\textbf{15.01}} & 48.68  & \underline{\textbf{15.01}}            \\
                     & SMAPE  & 33.70               & 61.60 & 160.40 & 76.20  & 39.50        & 41.50  & 44.20                 & 38.70 & 39.70 & {\color{black}40.26} & 62.50  & \underline{\textbf{28.30}} & 109.50 & \underline{\textbf{28.30}}            \\
\hline     
\end{tabular}
}
\label{Tab_NO2_30_Eval}
\end{table} 

The 60-day and 90-day-ahead forecasting results, as presented in Tables \ref{Tab_60_Eval}, \ref{Tab_60_Eval_NO2}, and \ref{Tab_90_Eval}, demonstrate how the proposed E-STGCN architecture improves upon the baseline models for longer forecast horizons. For both $\PM_{2.5}$ and $\PM_{10}$ pollutants, our model delivers the most accurate forecasts and quantifies the extreme observations during the first two 60-day windows, improving forecast accuracy by 9.73\% over the best-performing baseline model. In the subsequent two forecast periods (May–June and July–August), the GSTAR model performs best for $\PM_{2.5}$ levels, while for $\PM_{10}$, the proposed model provides similar average performance as the GpGp and NBeats frameworks and outperforms others in capturing the extreme patterns in the dataset. During the September–October period, the E-STGCN framework achieves the lowest forecast error for both pollutants. However, in the final medium-term forecast period of 2023, the ARIMA and the E-STGCN model surpass the performance of all other approaches in terms of point-based and distribution-based measures, respectively. For the medium-term and long-term forecasting of $\NO_2$ concentration levels, we observe similar patterns to those seen in the short-term forecasts. The proposed E-STGCN and modified STGCN models generate similar results and outperform the baseline models in most periods, except for July–August (in the medium-term) and the last two long-term forecast windows, where NBeats, DeepAR, and ARIMA perform better. For other forecast windows, the STARMA and GSTAR models offer competitive performance compared to the best-performing frameworks. For the long-term forecasting task of $\PM_{2.5}$, the E-STGCN architecture performs well during the first and last quarters of 2023, when the data predominantly exhibits more extreme behavior. During the summer months, the NBeats and DeepAR frameworks provide more accurate forecasts of $\PM_{2.5}$ concentration levels. In case of long-term forecasting of $\PM_{10}$, the proposed framework outperforms the competing models in capturing overall concentration trends during the first two and the last quarters of 2023. However, during the third forecast window (July–September), DeepAR, ARIMA, and NBeats achieve superior performance for forecasting $\PM_{10}$ levels. 

The experimental results reported in our study align with the \textit{No Free Lunch} theorem, which suggests that any forecasting model performing best on a particular dataset is likely to perform poorly on others \citep{wolpert1997no}. Overall, the E-STGCN framework consistently achieved superior forecast performance across most tasks. Among the temporal models, ARIMA and NBeats performed well, while from the spatiotemporal paradigm, most of the baseline architectures, namely STARMA, GSTAR, GpGp, STNN, and STGCN, demonstrated competitive performance. The DeepKriging framework, however, performed poorly in most forecasting tasks due to scalability issues, which hindered its ability to handle medium-sized spatiotemporal datasets. Additionally, the performance of models like LSTM, TCN, DeepAR, and Transformers lagged behind the E-STGCN framework due to their inability to effectively capture the spatial dependencies associated with pollutant concentrations. We also observed that the proposed E-STGCN consistently outperformed or performed similarly to the modified STGCN model. This advantage is attributed to the training mechanism adopted in E-STGCN, which employs the POT-based loss function. This modified loss function enables the framework to better capture exceedances of pollution concentration levels over the NAAQS threshold, especially during the onset and end of winter months. The performance of the E-STGCN model in the experimental results highlights its strong generalization ability and adaptability, making it capable of providing high forecast accuracy for datasets having extreme observations. Consequently, the E-STGCN framework offers effective and reliable forecasts for air pollutant concentrations across different horizons.

Additionally, we evaluated the computational feasibility of the E-STGCN model and baseline forecasters by recording the average training and inference time for each framework. Table \ref{table_computational_time} reports the total time (in seconds) required to train the models and generate 90-day ahead (January-March) forecasts for $PM_{2.5}$ concentrations across all monitoring stations of Delhi using a single core of an Intel i7 12th generation processor. Among the statistical approaches, ARIMA and STARMA require the least training time due to their low parameter complexity. In contrast, temporal-only deep learning models, particularly the Transformer and NBeats architectures, incur higher training durations due to their attention mechanisms and deep neural architectures, respectively. For spatiotemporal models, STGCN and modified STGCN require moderate training time, while the computational intensity of the GpGp framework increases its training duration. The proposed E-STGCN requires comparable training time to that of standard STGCN, despite the inclusion of an LSTM network in the temporal module and EVT-based loss function. While training times vary across models, inference remains fast across different frameworks except for GpGp and DeepKriging. This runtime analysis underscores that while E-STGCN introduces some additional computational demands, it remains competitive in terms of both training and inference time, and thus is well-suited for near real-time forecasting applications.
\begin{table}[!ht]
    \centering
    \footnotesize
    \caption{Forecasting performance of the proposed E-STGCN model in comparison to the temporal-only and spatiotemporal forecasting techniques for 60 60-day ahead forecast horizon of $PM_{2.5}$ and $PM_{10}$ pollutants (best results are \underline{\textbf{highlighted}}).}
    \resizebox{\textwidth}{!}{
\begin{tabular}{|c|c|c|cccccc|ccccccc|c|}
\hline \multirow{2}{*}{Pollutant} & Forecast & \multirow{2}{*}{Metric} & \multicolumn{6}{c|}{Temporal-only Model}                  & \multicolumn{7}{c|}{Spatiotemporal Model}           & Proposed \\ \cline{4-16}
                                 & \multirow{2}{*}{Period} & & \multirow{2}{*}{ARIMA} &  \multirow{2}{*}{LSTM} & \multirow{2}{*}{TCN} & \multirow{2}{*}{DeepAR} & \multirow{2}{*}{Transformers} & \multirow{2}{*}{NBeats} & \multirow{2}{*}{STARMA} & \multirow{2}{*}{GSTAR} & \multirow{2}{*}{GpGp} & \multirow{2}{*}{{\color{black}STGCN}} & \multirow{2}{*}{STNN} & Modified & \multirow{2}{*}{DeepKrigging} & \multirow{2}{*}{E-STGCN} \\ 
    & & & &  &  &  &  &  &  &  &  &  &  &  STGCN &  & \\ \hline

\multirow{36}{*}{PM$_{2.5}$} & \multirow{6}{*}{JAN - FEB} & MAE    & 73.48               & 132.20 & 116.87 & 131.90 & 66.52        & 59.28  & 80.56                 & 100.35 & 58.53 & {\color{black} 62.00}  & 76.26  & 56.73  & 84.47  & \underline{\textbf{55.00}}            \\
                       &                      & MASE   & 1.72                & 3.11   & 2.73   & 3.10   & 1.54         & 1.39   & 1.87                  & 2.36   & 1.36  & {\color{black} 1.44} & 1.82   & 1.32   & 1.99   & \underline{\textbf{1.28}}             \\
                       &                      & RMSE   & 86.23               & 148.60 & 135.33 & 148.33 & 89.53        & 77.12  & 96.46 & 113.61 & 78.81 & {\color{black} 84.09} & 94.98  & 76.44  & 100.93 & \underline{\textbf{73.00}}            \\
                       &                      & SMAPE  & 46.50               & 152.00 & 126.90 & 151.30 & 47.40        & 45.20  & 81.50                 & 57.00  & 39.90 & {\color{black} 43.58} & 48.80  & 39.60  & 77.20  & \underline{\textbf{37.90}}            \\
\hline
\end{tabular}
}
\label{Tab_60_Eval}
\end{table}

\begin{table}[!ht]
    \centering
    \footnotesize
    \caption{Forecasting performance of the proposed E-STGCN model in comparison to the temporal-only and spatiotemporal forecasting techniques for 60 60-day ahead forecast horizon of $NO_2$ pollutant (best results are \underline{\textbf{highlighted}}).}
    \resizebox{\textwidth}{!}{
\begin{tabular}{|c|c|c|cccccc|ccccccc|c|}
\hline \multirow{2}{*}{Pollutant} & Forecast & \multirow{2}{*}{Metric} & \multicolumn{6}{c|}{Temporal-only Model}                  & \multicolumn{7}{c|}{Spatiotemporal Model}           & Proposed \\ \cline{4-16}
                                 & \multirow{2}{*}{Period} & & \multirow{2}{*}{ARIMA} &  \multirow{2}{*}{LSTM} & \multirow{2}{*}{TCN} & \multirow{2}{*}{DeepAR} & \multirow{2}{*}{Transformers} & \multirow{2}{*}{NBeats} & \multirow{2}{*}{STARMA} & \multirow{2}{*}{GSTAR} & \multirow{2}{*}{GpGp} & \multirow{2}{*}{{\color{black}STGCN}} & \multirow{2}{*}{STNN} & Modified & \multirow{2}{*}{DeepKrigging} & \multirow{2}{*}{E-STGCN} \\ 
    & & & &  &  &  &  &  &  &  &  &  &  &  STGCN &  & \\ \hline
                    
\multirow{36}{*}{NO$_{2}$}  & \multirow{6}{*}{JAN - FEB} & MAE    & 15.04               & 29.56  & 39.64  & 26.76  & 17.21        & 18.62  & 18.45                 & 18.37  & 18.52 & {\color{black} 17.63} & 35.17  & \underline{\textbf{13.50}}  & 42.27  & \underline{\textbf{13.50}}            \\
                       &                      & MASE   & 2.85                & 6.53   & 9.23   & 5.79   & 5.78         & 4.21   & 3.85                  & \underline{\textbf{2.60}}   & 3.95 & {\color{black} 8.79}  & 7.72   & 4.69   & 8.48   & 4.69             \\
                       &                      & RMSE   & 18.17               & 32.57  & 42.33  & 29.93  & 20.18        & 22.76  & 22.59                 & 22.23  & 21.72 & {\color{black} 20.62} & 47.45  & \underline{\textbf{16.28}}  & 52.00  & \underline{\textbf{16.28}}            \\
                       &                      & SMAPE  & 37.30               & 91.50  & 172.00 & 76.70  & 41.50        & 51.50  & 46.50                 & 47.90  & 45.90 & {\color{black} 40.49} & 67.30  & \underline{\textbf{31.90}}  & 108.70 & \underline{\textbf{31.90}}            \\
\hline   
\end{tabular}
}
\label{Tab_60_Eval_NO2}
\end{table}

\begin{table}[!ht]
    \centering
    \footnotesize
    \caption{Forecasting performance of the proposed E-STGCN model in comparison to the temporal-only and spatiotemporal forecasting techniques for 90 90-day ahead forecast horizon of different pollutants (best results are \underline{\textbf{highlighted}}).}
    \resizebox{\textwidth}{!}{
\begin{tabular}{|c|c|c|cccccc|ccccccc|c|}
\hline \multirow{2}{*}{Pollutant} & Forecast & \multirow{2}{*}{Metric} & \multicolumn{6}{c|}{Temporal-only Model}                  & \multicolumn{7}{c|}{Spatiotemporal Model}           & Proposed \\ \cline{4-16}
                                 & \multirow{2}{*}{Period} & & \multirow{2}{*}{ARIMA} &  \multirow{2}{*}{LSTM} & \multirow{2}{*}{TCN} & \multirow{2}{*}{DeepAR} & \multirow{2}{*}{Transformers} & \multirow{2}{*}{NBeats} & \multirow{2}{*}{STARMA} & \multirow{2}{*}{GSTAR} & \multirow{2}{*}{GpGp} & \multirow{2}{*}{{\color{black}STGCN}} & \multirow{2}{*}{STNN} & Modified & \multirow{2}{*}{DeepKrigging} & \multirow{2}{*}{E-STGCN} \\ 
    & & & &  &  &  &  &  &  &  &  &  &  &  STGCN &  & \\ \hline 
                
\multirow{24}{*}{PM$_{2.5}$} & \multirow{6}{*}{JAN - FEB - MAR} & MAE    & 87.20               & 108.92 & 101.25 & 108.36 & 52.24        & 50.23  & 77.69                 & 80.22  & 50.66 & {\color{black} 51.90} & 91.36  & 47.42  & 65.84  & \underline{\textbf{46.00}}          \\
                       &                                  & MASE   & 2.47                & 3.10   & 2.87   & 3.09   & 1.48         & 1.43   & 2.20                  & 2.28   & 1.45   & {\color{black} 1.47} & 2.66   & 1.35   & 1.94   & \underline{\textbf{1.31}}             \\
                       &                                  & RMSE   & 97.95               & 127.48 & 118.65 & 127.00 & 75.09        & 66.30  & 90.16                 & 92.99  & 68.84  & {\color{black} 73.24} & 110.19 & 64.49  & 76.98  & \underline{\textbf{62.59}}          \\
                       &                                  & SMAPE  & 59.70               & 143.60 & 137.90 & 142.00 & 41.70        & 47.90  & 110.70                & 110.30 & 40.20  & {\color{black} 41.39} & 67.80  & 37.90  & 50.20  & \underline{\textbf{36.60}}           \\
\hline
\end{tabular}
}
\label{Tab_90_Eval}
\end{table}

\begin{table}[!ht]
    \centering
    \footnotesize
    \color{black}\caption{\color{black}Training and inference time (in seconds) for forecasting 90-day $PM_{2.5}$ concentration levels across all monitoring stations of Delhi.}
    \resizebox{\textwidth}{!}{
\begin{tabular}{|c|cccccc|ccccccc|c|}
\hline \multirow{3}{*}{Time} & \multicolumn{6}{c|}{Temporal-only Model} & \multicolumn{7}{c|}{Spatiotemporal Model}           & Proposed \\ \cline{2-15}
    & \multirow{2}{*}{ARIMA} &  \multirow{2}{*}{LSTM} & \multirow{2}{*}{TCN} & \multirow{2}{*}{DeepAR} & \multirow{2}{*}{Transformers} & \multirow{2}{*}{NBeats} & \multirow{2}{*}{STARMA} & \multirow{2}{*}{GSTAR} & \multirow{2}{*}{GpGp} & \multirow{2}{*}{STGCN} & \multirow{2}{*}{STNN} & Modified & \multirow{2}{*}{DeepKrigging} & \multirow{2}{*}{E-STGCN} \\ 
    & &  &  &  &  &  &  &  &  &  &  &  STGCN &  & \\ \hline
    Training & 0.33 & 29.85 & 28.82 & 29.37 & 181.75 & 544.54 & 0.16 & 2.82 & 285.95 & 170.01 & 0.18 & 155.02 & 116.95 & 176.83 \\
    Inference & 0.01 & 0.13 & 0.10 & 0.12 & 0.21 & 0.32 & 0.05 & 0.34 & 13.59 & 0.01 & 0.02 & 0.57 & 11.93 & 0.65\\  \hline
\end{tabular}}
\label{table_computational_time}
\end{table}

\subsection{Statistical Tests for Model Robustness}\label{Sec_statistical_significance}

\begin{figure}[!ht]
    \centering
    \includegraphics[width = 1\textwidth]{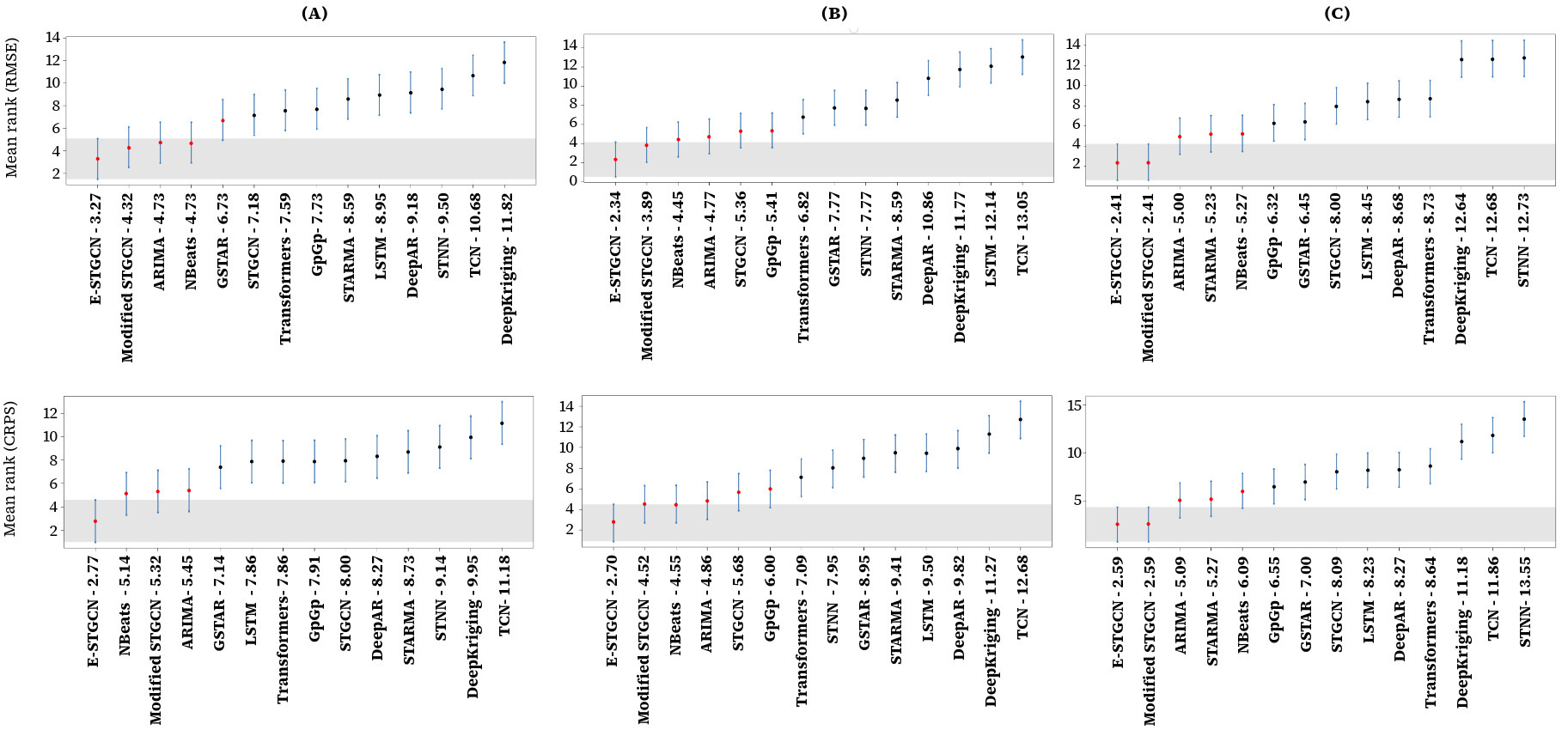}
    \caption{MCB Test results for (A) $\PM_{2.5}$, (B) $\PM_{10}$, and (C) $\NO_{2}$ pollutant concentration levels based on RMSE (upper panel) and CRPS (lower panel) metric. In the figure, for example, `E-STGCN-3.27' means that the average rank of the proposed E-STGCN algorithm based on the RMSE error metric is 3.27 for the $\PM_{2.5}$ dataset; the same explanation applies to other algorithms and datasets. The shaded region depicts the reference value of the test.}
    \label{Fig_MCB_RMSE}
\end{figure}

To validate the robustness of our experimental results, we employ multiple comparison with the best (MCB) test \citep{koning2005m3} and the Diebold-Mariano test \citep{diebold2002comparing}. The MCB test aims to identify the `best' forecasting model among all $\mathcal{F}$ architectures based on their performance across $\mathcal{D}$ datasets. For a specific evaluation metric, this non-parametric procedure ranks all models based on their performance across different forecasting tasks and computes the mean rank. The model with the lowest mean rank is considered the `best' forecasting architecture. Next, the critical distance (CD) for each of the $\mathcal{F}$ models is computed as $\delta_{\theta} \sqrt{\mathcal{F}(\mathcal{F} + 1)/6\mathcal{D}}$, where $\delta_{\theta}$ represents the critical value of the Tukey distribution at significance level $\theta$. {\color{black}The CD of the `best' performing model serves as the reference value against which all other models are compared. We apply the MCB test and visualize the results based on the deterministic RMSE and distribution-based CRPS metric for $\PM_{2.5}$, $\PM_{10}$, and $\NO_{2}$ in Fig. \ref{Fig_MCB_RMSE}. From the MCB plots (based on RMSE), we observe that the proposed E-STGCN architecture achieves the `best' performance, with a minimum rank of 3.27 for $\PM_{2.5}$ and 2.34 for $\PM_{10}$ datasets. The MCB test results based on the probabilistic CRPS metric showcase that the E-STGCN framework outperforms other models in capturing the overall patterns in the dataset, hence achieving the least rank of 2.77 for $\PM_{2.5}$ and 2.70 for $\PM_{10}$ datasets. For $\NO_{2}$ forecasting, the performance of E-STGCN and modified STGCN is similar, and they jointly obtain the lowest rank of 2.41 (in terms of RMSE) and 2.59 (in terms of CRPS).} Among the competing models, the modified STGCN, NBeats, and ARIMA frameworks consistently showcase better performance, having competitive ranks with E-STGCN. Spatiotemporal models such as GSTAR, STGCN, STARMA, and GpGp outperform the majority of the time-dependent frameworks by effectively capturing space-time correlations. Moreover, the CD values for most of the baseline models lie above the reference value (shaded region), indicating that their performance is significantly worse than the `best-fitted' E-STGCN model. Additional MCB test results based on the other evaluation metrics are provided in Fig. A.1, in Appendix A.3. {\color{black}Overall, the MCB test results, evaluated using both average and distributional forecast metrics, highlight that the proposed E-STGCN approach consistently produces accurate forecasts and effectively captures extreme concentration levels across various air pollutants.}

\begin{figure}[!ht]
    \centering
    \includegraphics[width = \textwidth]{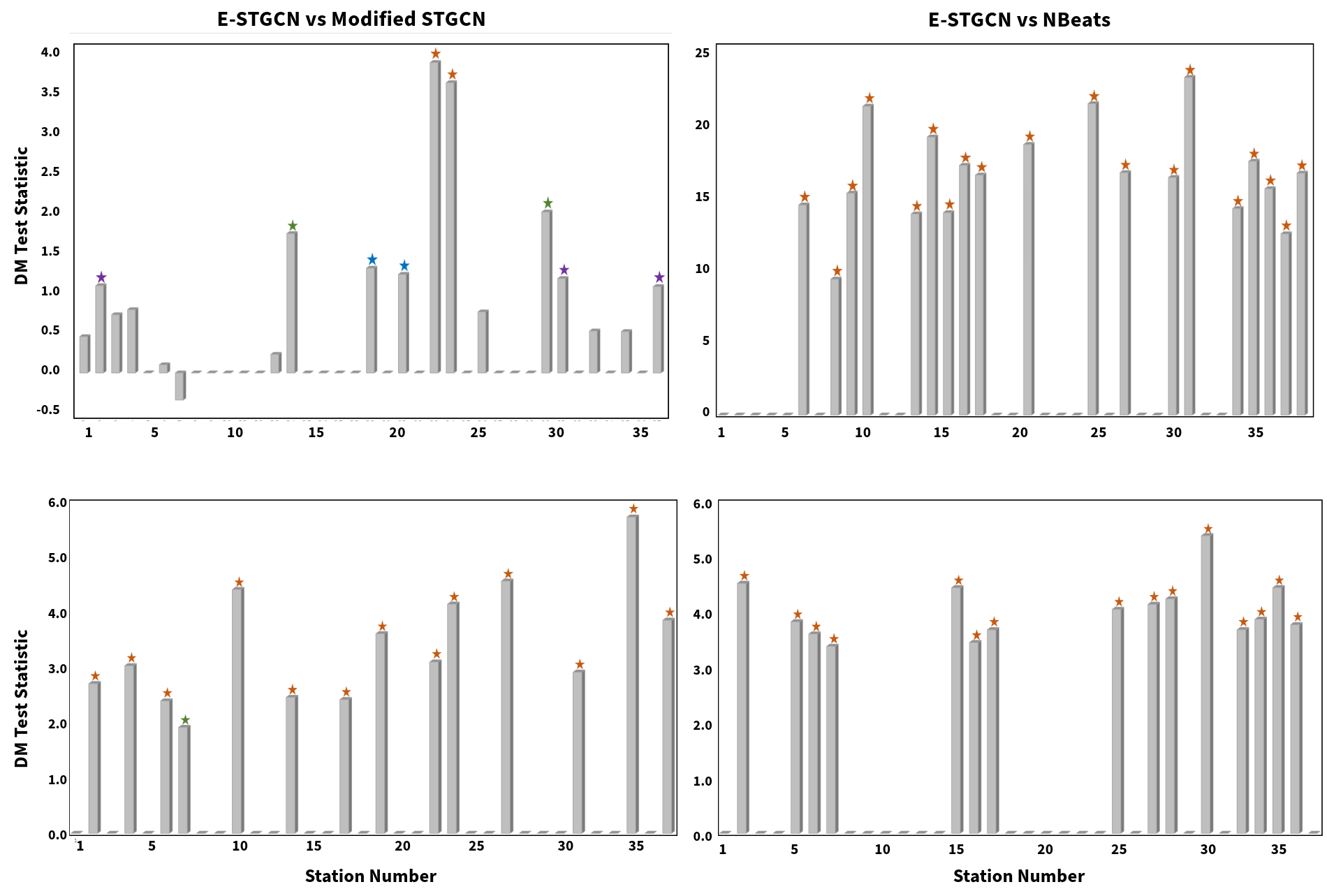}
    \caption{DM test results comparing (left) E-STGCN and Modified STGCN, and (right) E-STGCN and NBeats for forecasting $\PM_{2.5}$ (upper panel) and $\PM_{10}$ (lower panel) pollutant concentrations over the 90-day OCT-NOV-DEC forecast window. The Y-axis represents DM-test statistic values based on the MAE metric, while the X-axis indicates the monitoring station indices. Stars denote significant p-values, with colors representing 1\% (orange), 5\% (green), 10\% (blue), and 20\% (violet) significance levels, respectively.}
    \label{DM_Test_Pm25}
\end{figure}

Next, we employ the Diebold-Mariano (DM) test to assess whether the forecasting performance of the proposed E-STGCN framework significantly differs from that of the baseline models. Specifically, for any baseline architecture $\mathcal{A}$ and the proposed E-STGCN model, we compute the multivariate loss differential series for a given station as: 
\begin{equation*}
    \Lambda_{t, \mathcal{ A}}^i = \left | X_t^i - \widehat{X}_{t,\mathcal{ A}}^i\right | - \left |X_t^i - \widehat{X}_{t,\text{E-STGCN}}^i \right |,
\end{equation*}
where $X_t^i$ represents the ground truth data for station $i$ at time $t$ with $\widehat{X}_{t,\text{E-STGCN}}^i$ and $\widehat{X}_{t,\mathcal{ A}}^i$ being the corresponding forecasts generated by the E-STGCN and model $\mathcal{A}$, respectively. This statistical testing procedure checks whether the expected loss differential is zero using the DM statistic as 
$$\text{DM statistic for station } i = \sqrt{q} \frac{\mu_{\Lambda}^i}{\alpha_{\Lambda}^i},$$
where $q$ is the forecast horizon, $\mu_{\Lambda}^i$ and $\alpha_{\Lambda}^i$ are respectively the sample mean and standard deviation of the loss differential series $\Lambda_{t, \mathcal{ A}}^i$. Using this statistic, we test the null hypothesis $H_0 : \operatorname{\mathbb{E}}(\Lambda_{t, \mathcal{ A}}^i) \leqslant 0$ against the alternative $H_1 : \operatorname{\mathbb{E}}(\Lambda_{t, \mathcal{ A}}^i) > 0$, where $\operatorname{\mathbb{E}(\cdot)}$ denotes expectation. If the p-value of the test is less than the significance level, we reject $H_0$ and conclude that the forecasting performance of the E-STGCN framework is superior to that of $\mathcal{A}$ architecture. In our analysis, we conduct the DM test to assess the statistical significance of the performance differences between E-STGCN and the second and third-best-performing baselines, modified STGCN and NBeats. Fig.\ \ref{DM_Test_Pm25} present the test results for forecasting $\PM_{2.5}$ and $\PM_{10}$ concentrations during the October-November-December period. This plot evaluates the station-wise forecasting performance of E-STGCN with the benchmarks, where the x-axis represents station indices and the y-axis indicates DM test statistics. A positive DM test statistic value indicates the superiority of the E-STGCN over the baselines, while a negative value suggests that the baselines perform better. 

As highlighted in the plot, the E-STGCN method performs similarly to or better than the baselines across most stations, except for $\PM_{2.5}$ forecasting of CRRI Mathura Road (station no.\ 7), where modified STGCN achieves superior results. Moreover, the significant p-values at 1\%, 5\%, 10\%, and 20\% levels are marked using orange, green, blue, and violet-colored stars, respectively. As evident from Fig.\ \ref{DM_Test_Pm25}, E-STGCN significantly outperforms NBeats for $\PM_{2.5}$ forecasting in 19 out of 37 stations at a 1\% significance level. Compared to modified STGCN, E-STGCN demonstrates significantly different performance for multiple stations at varied levels. For the $\PM_{10}$ forecasting, we observe that E-STGCN achieves significantly better results than both the modified STGCN and NBeats for several monitoring stations at 1\% and 5\% levels of significance. The overall findings of the DM test are consistent with the MAE metrics reported in the experimental evaluations. Hence, this test underscores the statistical significance of our findings. For $\NO_{2}$ concentration levels, the forecasts from the E-STGCN and the modified STGCN models are very similar due to the absence of many significant extreme observations, resulting in $\Lambda_{t, \text{STGCN}}^i \approx 0$, rendering the DM statistic undefined in this case.

\subsection{Uncertainty Quantification}\label{Sec_uncertainty_quantification}

In addition to producing the point forecasts of the air pollutant concentrations through the E-STGCN approach, we quantify the uncertainty inherent with these forecasts using the conformal prediction technique \citep{vovk2005algorithmic}. This distribution-free approach generates the probabilistic intervals around the point estimates based on a conformal score $\left(\gamma_t\right)$. The computation of $\gamma_t$ at time $t$ involves modeling $p$-lagged values of the target series $\mathbf{X}_t$ using both E-STGCN and an uncertainty model $\mathcal{U}$ as follows
\begin{equation*}
    \gamma_t = \frac{|\mathbf{X}_{t} - \operatorname{E-STGCN}\left(\mathbf{X}_{t-p}\right)|}{\mathcal{U}\left(\mathbf{X}_{t-p}\right)}.
\end{equation*}

Subsequently, using the sequential nature of $\mathbf{X}_{t}$ and $\gamma_{t}$, we derive the conformal quantile by applying a weighted aggregation technique with a fixed window $\left\{\nu_t = \mathbbm{1}\left(\chi \geqslant t - \upsilon \right), \; \chi < t\right\}$ of size $\upsilon$ as
\begin{equation*}
    \kappa_t = \operatorname{inf}\left\{\omega : \frac{1}{\min \left(\upsilon, \chi - 1\right)} \sum_{\chi = 1}^{t-1} \gamma_{\chi} \nu_{\chi} \geqslant 1 - \rho \right\},
\end{equation*}
where $\rho$ is the significance level. Then, the computation of the conformal prediction interval using the conformal quantiles $\kappa_t$ can be expressed as,
\begin{equation*}
    \left[\operatorname{E-STGCN}\left(\mathbf{X}_{t-p}\right) \pm \kappa_t \; \mathcal{U}\left(\mathbf{X}_{t-p}\right) \right].
\end{equation*}

\begin{figure}[!ht]
    \centering
    \includegraphics[width = 0.8\textwidth]{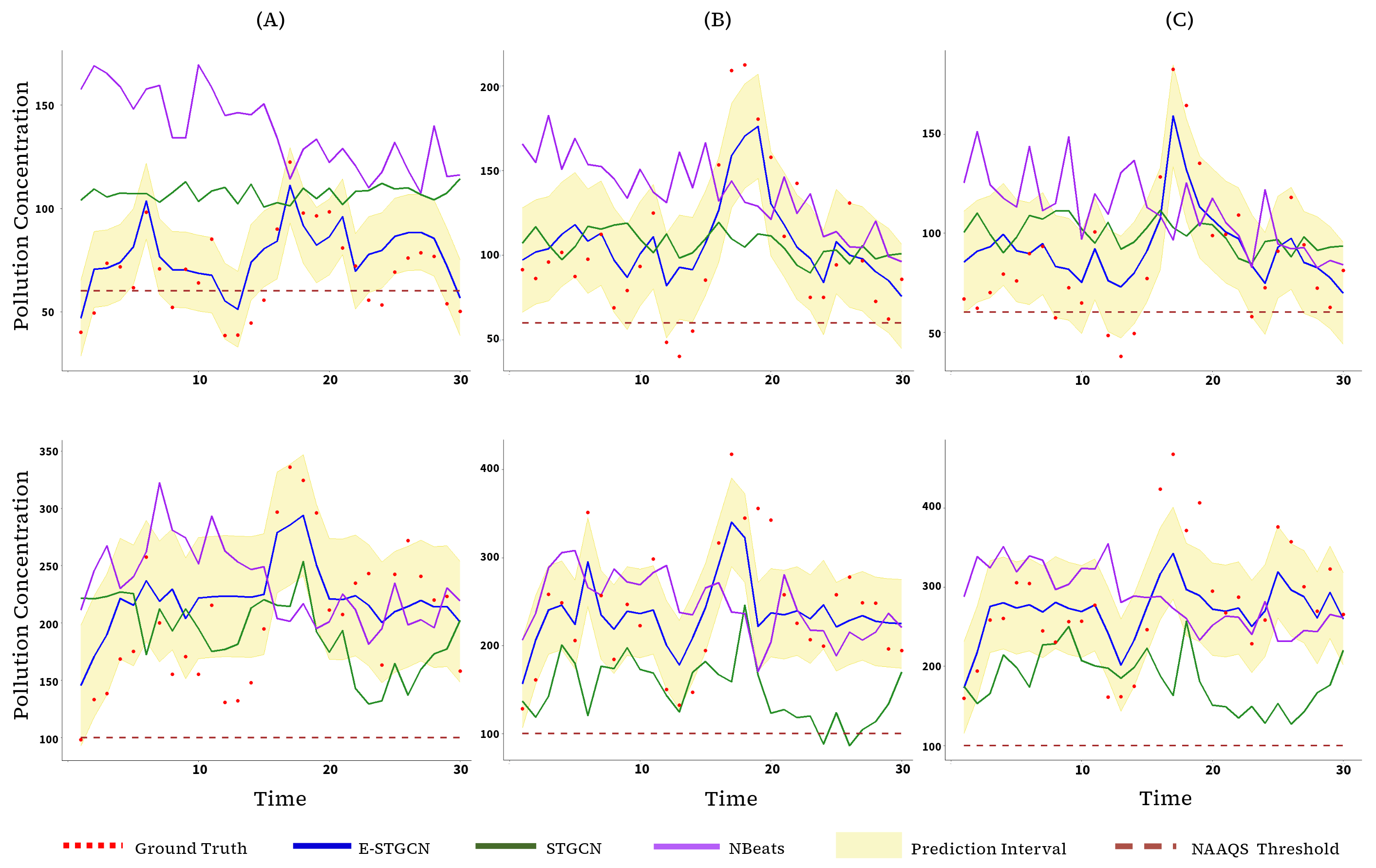}
    \caption{Upper panel presents ground truth (red dots) $\PM_{2.5}$ pollutant concentrations monitored at (A) DTU, (B) Dr. Karni Singh Shooting Range, and (C) IGI Airport stations during February 2023 window and corresponding point forecasts of E-STGCN (blue line), modified STGCN (green line), and NBeats (violet line) framework. The conformal prediction interval (yellow-shaded) of the E-STGCN model quantifies the associated uncertainty. The lower panel highlights similar information about $\PM_{10}$ concentrations monitored at the corresponding stations.}
    \label{CP_Plot_Fig}
\end{figure}

In Fig.\ \ref{CP_Plot_Fig}, we present the point and interval estimate of air pollutant concentrations generated by the E-STGCN model along with the results of the two best-performing models, modified STGCN and NBeats, as identified by the MCB plots in Section \ref{Sec_statistical_significance}. The conformal prediction intervals demonstrated in Fig.\ \ref{CP_Plot_Fig} are calculated with $\rho = 0.20$ for three selected forecasting scenarios. The plot highlights the generalizability of the proposed E-STGCN model in providing valuable insights about air pollutant concentration levels, mainly modeling their threshold exceedance values. These findings are pivotal for environmentalists in designing awareness and mitigation strategies.

\section{Policy implications}\label{Sec_Discussion} 

Rapid urbanization and industrialization have significantly impacted air quality in many developing and underdeveloped countries. In its 2021 Global Air Quality Guidelines (AQGs), WHO recommended critical air pollutants such as $\PM$, $\NO_2$, $\mathrm{SO}_2$, $\mathrm{O}_3$, and $\mathrm{CO}$ based on their effects on mortality and human health. Among these, $\PM$ and $\NO_2$ have gained particular attention from air quality researchers due to their direct links to increased mortality, as evidenced in epidemiological studies \citep{olaniyan2020association}. $\NO_2$ is a highly reactive gas primarily emitted from automobile exhaust, power plants, and industrial machinery. In urban areas, $\NO_2$ levels are mainly driven by the transportation sector. For instance, the urban regions of North America and Europe often report higher $\NO_2$ despite low levels of $\PM_{2.5}$ and $\PM_{10}$ \citep{ji2022no2}. It was found that acute exposure to $\NO_2$ can aggravate respiratory diseases, such as asthma and other pulmonary symptoms, although no causal relationship between $\NO_2$ exposure and health mortality was established \citep{faustini2014nitrogen}. On the other hand, $\PM$ comprises a mix of acids (such as nitrates and sulfates), organic chemicals, metals, soil, dust particles, and allergens. These particles originate from various sources like fossil fuel combustion, industrial emissions, construction activities, wildfires, stubble burning, and household cooking. In Delhi, $\PM$ levels frequently exceed the NAAQS, even when $\NO_2$ concentrations remain relatively low \citep{abirami2021regional}. It is particularly critical as $\PM$ is identified as a causal factor for cardiovascular and respiratory mortality and remains a serious concern for India’s capital. Given that the population of Delhi and the national capital region (NCR) of India is particularly vulnerable to $\PM$ exposure, which can cause health emergencies, this study aims to forecast $\PM$ levels in Delhi by considering extreme behaviors of air pollutants. The proposed E-STGCN framework offers a technological solution for real-time monitoring and forecasting of hazardous air pollutants in Delhi. This approach is particularly valuable when extreme observations and nonlinear patterns characterize the observed spatiotemporal data. The proposed methodology has the potential to advance future research endeavors on enhancing air quality forecasting models and to promote environmental sustainability. Although E-STGCN has been developed specifically for air pollution data in this study, it can also be extended to other applied fields, including epidemiology, seismology, and transportation research, where similar patterns of extreme events and complex dependencies are frequently observed.

\section{Conclusion}\label{Sec_Conclu}

Accurate air quality forecasting remains a challenging problem due to complex spatiotemporal dependencies in pollutant concentration levels. Pollutants such as $\PM_{2.5}$, $\PM_{10}$, and $\NO_{2}$ often exhibit extreme behaviors while also displaying nonlinear and non-stationary properties. Among these hazardous pollutants, $\PM_{2.5}$ and $\PM_{10}$ concentrations are consistently high in some of the world’s majorly polluted cities, leading to serious health hazards and restricting economic growth. In particular, air pollution levels can intensify with seasonal variations. For instance, in Delhi, the concentration of $\PM_{2.5}$ and $\PM_{10}$ increases rapidly during winter due to low wind speed, stubble burning, firecracker emissions, and other contributing sources. To address these challenges, public awareness through early warning systems is of paramount importance. 

In this study, we propose the E-STGCN model, which aims to provide actionable insights by generating real-time forecasts of air pollutant concentrations. Our approach bridges the gap between existing EVT-based models, which focus on predicting extreme behavior, and data-driven forecasting methods that predict future trajectories without accounting for the tail behavior of the extreme observations. By integrating EVT knowledge with modified spatiotemporal GCNs, our proposed framework effectively performs spatiotemporal forecasting while tackling extreme observations. Experimental results, conducted on real-world air pollutant data (daily frequencies) of $\PM_{2.5}$, $\PM_{10}$, and $\NO_{2}$ from 37 monitoring stations in Delhi demonstrate that the E-STGCN approach is well-suited for predicting the future dynamics of nonlinear and non-stationary datasets with spatiotemporal dependence and extreme events. Additionally, the model generates appropriate probabilistic bands along with point forecasts, enabling environmental advocates to monitor air pollution trends and design effective control strategies. Further, the forecastability and statistical significance tests conducted in this study verify the effectiveness and robustness of the proposed architecture for air pollution forecasting over various time horizons. 

An interesting avenue for future research is to identify various climatic, transportation, and industrial indices that have a causal impact on rising air pollution levels. Future studies could explore how incorporating these causal covariates might enhance the accuracy of the E-STGCN approach. {\color{black}Additionally, assessing model performance on block maxima (e.g., monthly or seasonal peaks) could offer deeper insights into the model’s ability to forecast extreme events. Such an evaluation would complement the EVT-guided training objective by directly linking statistical tail modeling to empirical performance during high-impact pollution events.} Another potential direction would be to extend the air pollution modeling capabilities of E-STGCN on a global scale, analyzing its impact on pollution-related mortality and morbidity.

\section*{Competing interests}
No competing interest is declared.

\section*{Data and Code Availability Statement}
Data and codes are available in our GitHub repository: \url{https://github.com/mad-stat/E_STGCN}.

\bibliography{refs}


\newpage

\section*{Appendices}

\appendix

\renewcommand\thesection{A.\arabic{section}}
\renewcommand\thetable{A.\arabic{table}}
\renewcommand\thefigure{A.\arabic{figure}}

\setcounter{table}{0}
\setcounter{figure}{0}
\setcounter{section}{0}

\section{Statistical Tests on Air Pollutant Data}\label{App_Global_Feat}

Below, we summarize the global features of the dataset analyzed in our study and highlight their implementation strategies: 
\begin{itemize}
    \item \textit{Long-term dependency} is a crucial feature in time series processes and has gained significant attention in probabilistic time series modeling. To evaluate long-range dependency, we examine the self-similarity parameter, often referred to as the Hurst exponent, using the \textit{pracma} package in R.
    \item \textit{Stationarity} is a fundamental property of time series that implies the statistical features, particularly mean and variance, remain constant over time. To assess stationarity, we employ the Kwiatkowski–Phillips–Schmidt–Shin (KPSS) test from the \textit{tseries} package in R.
    \item \textit{Linearity} is another essential characteristic of time series data, playing a critical role in model selection. In our analysis, we apply Terasvirta’s neural network test from the \textit{nonlinearTseries} package in R to determine whether the data-generating process follows a linear trend.
    \item \textit{Seasonality} refers to recurring patterns in time series that occur at regular intervals. To identify the presence and frequency of seasonal patterns in our dataset, we use Ollech and Webel’s test from the \textit{seastests} package in R.
\end{itemize}

Descriptive statistics and these global features of the three pollutant series ($\PM_{2.5}$, $\PM_{10}$, and $\NO_2$) for various monitoring stations in our dataset are presented in Tables \ref{Sum_Stat_PM25}, \ref{Sum_Stat_PM10} and \ref{Sum_Stat_NO2}.

\begin{table}[!ht]
\centering
\tiny
\caption{Descriptive statistics of the PM$_{2.5}$ pollutant levels in different stations. In the table, weekly (W), quarterly (Q), and no (-) seasonality are indicated.}
\begin{tabular}{|lp{0.02\textwidth}p{0.06\textwidth}p{0.03\textwidth}p{0.035\textwidth}p{0.06\textwidth}p{0.03\textwidth}p{0.03\textwidth}p{0.03\textwidth}p{0.05\textwidth}p{0.04\textwidth}p{0.05\textwidth}p{0.03\textwidth}p{0.03\textwidth}|} 
\hline 
Station  & Min   & 1st quartile & Mean   & Median & 3rd quartile & Max    & Sd     & CV   & Skewness & Kurtosis & Freq of EVal & Seas & DW test \\
1 | Alipur  & 4.81  & 48.15        & 127.64 & 95.71  & 181.72       & 758.40  & 103.66 & 0.81 & 1.42     & 2.36     & 68.01        & -    & 0.12            \\
2 | Anand Vihar                         & 9.52  & 52.72        & 122.39 & 88.94  & 162.3        & 592.28 & 96.17  & 0.79 & 1.52     & 2.24     & 69.66        & W    & 0.11           \\
3 | Ashok Vihar                         & 5.95  & 42.35        & 109.27 & 75.6   & 146.49       & 601.49 & 93.26  & 0.85 & 1.68     & 3.13     & 60.41        & -    & 0.06           \\
4 | Aya Nagar                           & 7.96  & 37.22        & 77.86  & 58.01  & 98.12        & 556.03 & 61.17  & 0.79 & 2.33     & 8.45     & 47.95        & W    & 0.24           \\
5 | Bawana                              & 7.83  & 52.92        & 122.94 & 94.93  & 166.03       & 696.78 & 93.11  & 0.76 & 1.55     & 3.19     & 69.38        & -    & 0.05           \\
6 | Burari Crossing                     & 10.97 & 81.64        & 105.04 & 99.79  & 118.07       & 565.65 & 53.53  & 0.51 & 2.33     & 9.28     & 83.36        & -    & 0.37          \\
7 | CRRI Mathura Road & 7.97  & 42.72        & 97.18  & 72.76  & 125.91       & 518.18 & 76.23  & 0.78 & 1.78     & 3.86     & 59.52        & Q    & 0.24           \\
8 | DTU  & 9.39  & 46.63        & 111.06 & 83.60   & 149.51       & 631.85 & 86.41  & 0.78 & 1.67     & 3.65     & 65.27        & -    & 0.84           \\
9 | Dr. Karni Singh Shooting Range  & 4.39  & 34.27        & 89.47  & 60.34  & 122.61       & 571.73 & 79.41  & 0.89 & 1.79     & 3.89     & 50.21        & Q    & 0.09 \\
10 | Dwarka Sector 8                    & 8.09  & 41.38        & 104    & 76.03  & 140.48       & 600.4  & 85.35  & 0.82 & 1.72     & 3.88     & 59.59        & -    & 0.03           \\
11 | IGI Airport                        & 2.68  & 34.89        & 82.85  & 60.57  & 108.71       & 506.20  & 67.44  & 0.81 & 1.98     & 5.51     & 50.55        & -    & 0.01           \\
12 | Ihbas Dilshad Garden              & 7.50   & 43.27        & 91.83  & 76.73  & 119.43       & 606.41 & 66.98  & 0.73 & 1.70      & 4.67     & 62.40         & -    & 0.63          \\  
13 | ITO & 12.13 & 53.52        & 111.17 & 84.02  & 142.13       & 659.29 & 84.56  & 0.76 & 2.04     & 5.86     & 69.25        & -    & 0.29          \\
14 | Jahangirpuri                       & 8.77  & 49.88        & 128.39 & 90.71  & 179.66       & 658.26 & 105.42 & 0.82 & 1.46     & 2.03     & 67.67        & -    & 0.05           \\
15 | Jawaharlal Nehru Stadium & 3.79  & 36.18        & 95.16  & 65.22  & 128.78       & 513.36 & 82.48  & 0.87 & 1.69     & 3.17     & 53.84        & -    & 0.05           \\
16 | Lodhi Road IMD                     & 5.49  & 38.08        & 78.41  & 60.37  & 99.07        & 479.72 & 58.98  & 0.75 & 2.05     & 5.80      & 50.34        & -    & 0.26           \\
17 | Major Dhyan Chand National Stadium & 6.71  & 38.98        & 94.15  & 67.88  & 126.01       & 525.64 & 76.45  & 0.81 & 1.63     & 3.07     & 55.41        & W    & 0.75           \\
18 | Mandir Marg                        & 4.67  & 39.87        & 93.91  & 73.81  & 121.22       & 563.93 & 74.69  & 0.8  & 1.75     & 4.07     & 59.45        & -    & 0.18           \\
19 | Mundka                             & 7.38  & 45.92        & 120.93 & 93.04  & 169.61       & 698.96 & 98.22  & 0.81 & 1.55     & 3.12     & 65.27        & -    & 0.20           \\
20 | Najafgarh                          & 4.50   & 35.22        & 83.68  & 67.57  & 111.60        & 574.41 & 65.59  & 0.78 & 1.92     & 6.09     & 55.89        & Q    & 0.31 \\
21 | Narela                             & 6.62  & 47.04        & 110.17 & 84.39  & 149.57       & 689.10  & 85.87  & 0.78 & 1.62     & 3.68     & 65.62        & -    & 0.19          \\
22 | Nehru Nagar                        & 6.06  & 41.86        & 116.12 & 74.08  & 163.04       & 554.05 & 101.59 & 0.87 & 1.48     & 1.96     & 59.45        & W    & 0.03           \\
23 | North Campus DU                    & 5.58  & 43.00           & 95.97  & 70.35  & 123.72       & 570.07 & 77.70   & 0.81 & 1.87     & 4.43     & 58.22        & -    & 0.16           \\
24 | NSIT Dwarka                        & 8.33  & 51.39        & 100.39 & 86.92  & 133.17       & 406.61 & 65.08  & 0.65 & 1.23     & 1.72     & 68.42        & -    & 0.76            \\
25 | Okhla Phase 2                      & 6.18  & 37.48        & 99.64  & 67.85  & 132.97       & 547.56 & 87.48  & 0.88 & 1.72     & 3.22     & 55.07        & W    & 0.07           \\
26 | Patparganj                         & 4.73  & 42.34        & 104.57 & 74.03  & 138.85       & 633.95 & 87.84  & 0.84 & 1.65     & 3.21     & 60.41        & W    & 0.25            \\
27 | Punjabi Bagh & 6.84  & 46.34        & 109.45 & 79.57  & 144.22       & 609.18 & 88.64  & 0.81 & 1.84     & 4.40      & 62.67        & -    & 0.04           \\
28 | PUSA DPCC & 3.68  & 37.14        & 93.82  & 67.32  & 130.31       & 570.61 & 79.06  & 0.84 & 1.64     & 3.29     & 54.25        & W    & 0.11            \\
29 | PUSA IMD & 7.46  & 35.75        & 80.82  & 60.75  & 103.87       & 569.5  & 64.86  & 0.80  & 2.13     & 7.01     & 50.89        & -    & 0.06           \\
30 | R K Puram & 6.38  & 41.25        & 99.97  & 71.01  & 134.12       & 558.86 & 83.07  & 0.83 & 1.61     & 3.01     & 58.29        & -    & 0.06           \\
31 | Rohini & 6.13  & 47.64        & 119.07 & 85.15  & 160.09       & 761.95 & 99.09  & 0.83 & 1.72     & 3.81     & 64.79        & -    & 0.89          \\
32 | Shadipur                           & 9.38  & 37.88        & 89.67  & 72.40   & 119.80        & 397.45 & 66.43  & 0.74 & 1.40     & 2.13     & 58.42        & -    & 0.35           \\
33 | Sirifort                           & 0.08  & 39.96        & 96.32  & 70.71  & 129.24       & 573.09 & 78.07  & 0.81 & 1.79     & 4.38     & 57.95        & W    & 0.40           \\
34 | Sonia Vihar                        & 6.01     & 46.37        & 106.71 & 77.84  & 138.55       & 598.80  & 86.30   & 0.81 & 1.75     & 3.69     & 62.60         & -    & 0.34 \\
35 | Sri Aurobindo Marg & 5.28  & 34.88  & 85.33  & 60.77  & 113.41       & 534.86 & 72.01  & 0.84 & 1.94  & 5.31     & 50.68 & -    & 0.17           \\
36 | Vivek Vihar  & 4.61  & 45.91        & 115.35 & 79.91  & 156.20        & 650.88 & 97.79  & 0.85 & 1.57     & 2.48     & 62.60         & -    & 0.27  \\
37 | Wazirpur                           & 6.88  & 48.74        & 114.83 & 80.10   & 150.70        & 585.79 & 92.57  & 0.81 & 1.69     & 3.04     & 65.41 & -    & 0.15    \\ \hline     
\end{tabular}
\label{Sum_Stat_PM25}
\end{table}

\begin{table}[!ht]
\centering
\tiny
\caption{Descriptive statistics of the PM$_{10}$ pollutant levels in different stations. In the table, weekly (W), quarterly (Q), and no (-) seasonality are indicated.}
\begin{tabular}
{|lp{0.03\textwidth}p{0.06\textwidth}p{0.06\textwidth}p{0.035\textwidth}p{0.06\textwidth}p{0.03\textwidth}p{0.03\textwidth}p{0.03\textwidth}p{0.05\textwidth}p{0.04\textwidth}p{0.05\textwidth}p{0.05\textwidth}p{0.03\textwidth}|} 
\hline 
Station  & Min   & 1st quartile & Mean   & Median & 3rd quartile & Max    & Sd     & CV   & Skewness & Kurtosis & Freq of EVal & Seas & DW test \\ \hline
1       & 10.04 & 98.38        & 201.54 & 183.66 & 284.63       & 758.40  & 125.43 & 0.62 & 0.76     & 0.36     & 74.38        & -    & 0.20    \\
2       & 16.26 & 128.95       & 265.22 & 249.69 & 361.79       & 729.94 & 153.63 & 0.58 & 0.57     & -0.38    & 83.56        & -    & 0.31    \\
3       & 11.80  & 106.94       & 212.58 & 191    & 290.86       & 753.97 & 130.1  & 0.61 & 0.91     & 0.59     & 77.47        & -    & 0.16    \\
4       & 11.39 & 79.36        & 152.38 & 138.25 & 200.58       & 665.41 & 94.31  & 0.62 & 1.22     & 2.20      & 66.16        & W    & 0.17    \\
5       & 12.03 & 127.09       & 240.84 & 218.46 & 335.64       & 810.85 & 139.82 & 0.58 & 0.73     & 0.16     & 82.74        & -    & 0.20    \\
6       & 20.33 & 135.13       & 201.05 & 180.15 & 226.15       & 795.88 & 99.65  & 0.50  & 1.40     & 2.61     & 91.99        & -    & 1.00    \\
7       & 10.22 & 92.00 & 205.10  & 189.72 & 292.14       & 721.69 & 129.26 & 0.63 & 0.74     & 0.22     & 72.19        & -    & 0.45    \\
8       & 1.00     & 114.05       & 214.90  & 199.14 & 295.47       & 923.70  & 123.74 & 0.58 & 0.78     & 0.87     & 78.97        & -    & 0.72    \\
9       & 8.03  & 87.93        & 180.60  & 166.45 & 243.90        & 756.47 & 110.83 & 0.61 & 0.88     & 0.99     & 71.37        & Q, W & 0.52    \\
10      & 14.63 & 136.14       & 253.11 & 247.16 & 351.80        & 807.89 & 136.80  & 0.54 & 0.47     & -0.21    & 85.21        & -    & 0.38    \\
11      & 12.95 & 87.57        & 172.51 & 148.58 & 235.35       & 633.67 & 105.43 & 0.61 & 1.13     & 1.35     & 68.22        & W    & 0.01    \\
13      & 16.00    & 94.81        & 170.99 & 153.03 & 226.25       & 691.84 & 99.54  & 0.58 & 1.21     & 2.17     & 71.85        & -    & 0.75    \\
14      & 15.48 & 125.76       & 248.47 & 229.62 & 339.05       & 821.78 & 145.85 & 0.59 & 0.72     & 0.03     & 81.99        & -    & 0.23    \\
15      & 10.25 & 96.43        & 190.98 & 175.62 & 260.25       & 678.37 & 115.37 & 0.60  & 0.86     & 0.71     & 74.04        & Q, W & 0.46    \\
16      & 10.06 & 84.84        & 161.70  & 147.87 & 218.69       & 611.18 & 93.82  & 0.58 & 0.95     & 1.05     & 69.45        & -    & 0.18    \\
17      & 12.44 & 98.55        & 190.49 & 171.7  & 261.15       & 663.12 & 112.55 & 0.59 & 0.82     & 0.39     & 74.32        & -    & 0.32    \\
18      & 17.09 & 92.99        & 172.77 & 159.56 & 231.77       & 705.68 & 98.16  & 0.57 & 0.90      & 0.96     & 71.58        & -    & 0.46    \\
19      & 10.67 & 129.55       & 256.72 & 241.74 & 358.95       & 790.97 & 144.14 & 0.56 & 0.52     & -0.35    & 84.32        & -    & 0.54    \\
20      & 8.75  & 86.20         & 157.88 & 147.47 & 212.83       & 731.23 & 93.13  & 0.59 & 0.93     & 1.54     & 69.18        & Q    & 0.92    \\
21      & 20.48 & 128.72       & 231.14 & 207.18 & 314.71       & 718.18 & 127.53 & 0.55 & 0.75     & 0.16     & 85.07        & -    & 0.21    \\
22      & 10.48 & 97.90         & 205.37 & 181.25 & 284.89       & 702.19 & 129.16 & 0.63 & 0.90  & 0.53     & 74.11        & W    & 0.29    \\
23      & 5.17  & 100.15       & 194.37 & 176.32 & 260.85       & 735.53 & 119.71 & 0.62 & 1.00        & 1.10     & 75.07        & -    & 0.38    \\
25      & 9.73  & 108.34       & 211.05 & 190.43 & 282.79       & 741.37 & 124.62 & 0.59 & 0.86     & 0.56     & 78.08        & W    & 0.34    \\
26      & 8.36  & 90.98        & 189.19 & 170.67 & 262.95       & 689.97 & 118.8  & 0.63 & 0.83     & 0.36     & 71.85        & W    & 0.23    \\
27      & 20.60  & 109.55       & 206.63 & 183.98 & 277.36       & 768.33 & 121.4  & 0.59 & 0.96     & 0.79     & 78.36        & W    & 0.27    \\
28      & 9.97  & 104.87       & 201.78 & 191.35 & 277.41       & 726.86 & 117.88 & 0.58 & 0.66     & 0.26     & 76.64        & -    & 0.79    \\
29      & 13.01 & 69.98        & 152.34 & 130.26 & 209.25       & 706.66 & 99.95  & 0.66 & 1.24     & 2.02     & 62.12        & W    & 0.22    \\
30      & 11.52 & 98.04        & 193.74 & 180.34 & 266.95       & 699.3  & 111.82 & 0.58 & 0.74     & 0.46     & 74.45        & W    & 0.09    \\
31      & 11.1  & 116.87       & 230.09 & 204.46 & 318.69       & 783.4  & 138.92 & 0.60  & 0.82     & 0.21     & 80.14        & -    & 0.09    \\
33      & 10.97 & 118.90        & 221.00    & 213.68 & 301.66       & 664.34 & 121.23 & 0.55 & 0.63     & 0.18     & 82.05        & -    & 0.50    \\
34      & 13.00    & 108.97       & 213.34 & 190.00    & 292.41       & 720.11 & 127.22 & 0.60  & 0.88     & 0.50      & 78.77        & -    & 0.75    \\
35      & 8.86  & 72.39        & 148.27 & 134.51 & 202.46       & 596.06 & 92.57  & 0.62 & 1.00        & 1.23     & 63.15        & -    & 0.53    \\
36      & 12.48 & 111.68       & 223.87 & 198.14 & 305.24       & 699.36 & 132.67 & 0.59 & 0.85     & 0.32     & 79.18        & W    & 0.33    \\
37      & 19.3  & 138.93       & 246.48 & 215.07 & 325.32       & 793.65 & 139.44 & 0.57 & 0.99     & 0.68     & 88.42        & -    & 0.04  \\ \hline 
\end{tabular}
\label{Sum_Stat_PM10}
\end{table}

\begin{table}[!ht]
\centering
\tiny
\caption{Descriptive statistics of the NO$_{2}$ pollutant levels in different stations. In the table, weekly (W), quarterly (Q), and no (-) seasonality are indicated.}
\begin{tabular}
{|lp{0.03\textwidth}p{0.06\textwidth}p{0.06\textwidth}p{0.035\textwidth}p{0.06\textwidth}p{0.03\textwidth}p{0.03\textwidth}p{0.03\textwidth}p{0.05\textwidth}p{0.04\textwidth}p{0.05\textwidth}p{0.05\textwidth}p{0.03\textwidth}|} 
\hline 
Station  & Min   & 1st quartile & Mean   & Median & 3rd quartile & Max    & Sd     & CV   & Skewness & Kurtosis & Freq of EVal & Seas & DW test \\ \hline
1       & 0.70   & 17.88        & 34.28  & 30.93  & 46.07        & 120.01 & 20.60   & 0.60  & 0.99     & 1.00        & 3.70          & -    & 0.46    \\
2       & 4.87  & 44.92        & 75.56  & 70.53  & 100.18       & 325.77 & 39.95  & 0.53 & 1.06     & 2.55     & 41.03        & Q    & 0.25    \\
3       & 3.54  & 21.59        & 40.42  & 36.01  & 55.90         & 183.59 & 23.53  & 0.58 & 0.96     & 1.57     & 5.41         & -    & 0.83    \\
4       & 0.94  & 13.15        & 19.48  & 18.58  & 23.76        & 128.24 & 11.50   & 0.59 & 2.29     & 10.54    & 0.07         & -    & NA      \\
5       & 1.85  & 11.60         & 27.25  & 21.86  & 36.65        & 188.92 & 21.39  & 0.79 & 1.98     & 7.01     & 2.40          & -    & 0.23    \\
6       & 1.54  & 34.17        & 142.87 & 86.02  & 250.16       & 428.15 & 130.93 & 0.92 & 0.72     & -0.90     & 51.37        & Q    & 1.00       \\
7       & 0.33  & 14.00           & 38.52  & 20.72  & 41.88        & 308.38 & 43.89  & 1.14 & 2.30      & 5.01     & 12.67        & -    & 0.29    \\
8       & 1.06  & 21.43        & 40.66  & 33.89  & 47.72        & 276.46 & 32.42  & 0.8  & 2.41     & 7.97     & 8.77         & -    & 0.51    \\
9       & 3.44  & 28.22        & 53.46  & 46.16  & 72.41        & 291.46 & 34.51  & 0.65 & 1.64     & 6.14     & 20.07        & -    & 0.24    \\
10      & 6.61  & 21.27        & 36.5   & 31.97  & 47.19        & 127.53 & 20.34  & 0.56 & 1.16     & 1.40      & 3.70          & -    & 0.46    \\
11      & 0.64  & 23.9         & 44.08  & 33.03  & 56.03        & 417.31 & 37.91  & 0.86 & 3.43     & 21.22    & 11.58        & -    & 0.01    \\
12      & 6.32  & 26.16        & 50.41  & 42.99  & 67.89        & 197.44 & 31.23  & 0.62 & 1.17     & 1.48     & 16.71        & -    & 0.59    \\
13      & 8.13  & 19.87        & 36.33  & 28.14  & 40.49        & 272.12 & 29.42  & 0.81 & 3.23     & 14.19    & 6.99         & -    & 0.23    \\
14      & 8.63  & 31.49        & 60.03  & 50.28  & 73.94        & 237.59 & 39.34  & 0.66 & 1.62     & 2.63     & 20.96        & -    & 1.00       \\
15      & 6.51  & 40.86        & 60.95  & 58.68  & 78.80         & 202.93 & 27.71  & 0.45 & 0.55     & 0.54     & 23.90        & -    & 0.47    \\
16      & 0.13  & 6.49         & 13.73  & 9.99   & 18.13        & 100.24 & 11.54  & 0.84 & 2.15     & 7.10      & 0.21         & -    & 1.00       \\
17      & 8.97  & 23.66        & 42.06  & 35.99  & 53.79        & 159.29 & 24.17  & 0.57 & 1.25     & 1.68     & 8.70          & -    & 0.48    \\
18      & 11.15 & 39.88        & 54.12  & 52.51  & 69           & 179.24 & 22.53  & 0.42 & 0.46     & 0.64     & 12.26        & -    & 0.67    \\
19      & 4.38  & 23.26        & 37.74  & 33.95  & 49.39        & 117.4  & 18.32  & 0.49 & 0.85     & 0.47     & 2.33         & -    & 0.38    \\
20      & 2.96  & 11.74        & 20.73  & 18.23  & 27.27        & 89.38  & 12.29  & 0.59 & 1.36     & 2.90      & 0.14         & -    & NA      \\
21      & 4.12  & 26.27        & 38.89  & 34.96  & 48.52        & 150.4  & 17.65  & 0.45 & 1.21     & 2.25     & 3.22         & -    & 0.49    \\
22      & 8.63  & 34.86        & 55.21  & 48.80   & 71.95        & 225.00    & 27.77  & 0.50  & 1.30      & 3.23     & 17.53        & -    & 0.93    \\
23      & 0.96  & 9.09         & 25.18  & 17.37  & 30.10         & 205.36 & 24.75  & 0.98 & 2.34     & 6.74     & 5.27         & -    & 0.24    \\
24      & 2.09  & 19.09        & 29.37  & 26.55  & 37.99 & 101.78 & 14.01  & 0.48 & 1.20      & 2.49     & 0.89         & -    & 0.36    \\
25      & 7.12  & 29.30         & 49.55  & 45.01  & 65.23        & 219.36 & 25.99  & 0.52 & 0.96     & 1.80      & 14.59        & -    & 0.08    \\
26      & 1.78  & 15.05        & 31.51  & 24.23  & 38.31        & 140.9  & 24.21  & 0.77 & 1.59     & 2.44     & 5.75         & -    & 0.44    \\
27      & 0.22  & 33.69        & 51.52  & 46.93  & 63.70         & 208.14 & 25.11  & 0.49 & 1.58     & 5.26     & 11.10         & -    & 0.51    \\
28      & 6.57  & 33.97        & 53.48  & 54.06  & 71.61        & 155.5  & 24.43  & 0.46 & 0.14     & -0.63    & 14.59        & -    & 0.50     \\
29      & 0.65  & 12.32        & 35.21  & 23.84  & 47.35        & 325.08 & 35.86  & 1.02 & 2.57     & 10.75    & 9.86         & -    & 0.40     \\
30      & 0.27  & 27.91        & 44.34  & 43.06  & 59.15        & 179.95 & 22.45  & 0.51 & 0.53     & 1.06     & 6.71         & -    & 0.70     \\
31      & 0.30   & 14.48        & 25.59  & 21.53  & 32.62        & 146.91 & 15.69  & 0.61 & 1.72     & 5.22     & 0.75         & -    & 0.51    \\
32      & 6.37  & 24.22        & 51.60   & 44.04  & 73.03        & 181.41 & 32.76  & 0.63 & 0.95     & 0.58     & 19.18        & -    & 0.44    \\
33      & 0.21  & 14.10         & 34.24  & 25.96  & 48.55        & 247.18 & 27.86  & 0.81 & 1.73     & 5.02     & 7.05         & -    & 0.20     \\
34      & 2.87  & 21.53        & 35.90   & 32.32  & 46.62        & 111.92 & 18.41  & 0.51 & 0.91     & 0.77     & 2.74         & Q    & 0.42    \\
35      & 2.22  & 20.39        & 29.54  & 28.34  & 36.58        & 106.78 & 12.91  & 0.44 & 0.95     & 2.18     & 0.27         & -    & NA       \\
36      & 0.15  & 17.76        & 29.64  & 25.19  & 39.45        & 107.11 & 16.14  & 0.54 & 0.98     & 0.95     & 0.75         & -    & 0.67    \\
37      & 2.05  & 23.73        & 41.77  & 37.17  & 55.93        & 139.69 & 23.67  & 0.57 & 0.87     & 0.60     & 7.88         & -    & 0.55   \\ \hline
\end{tabular}
\label{Sum_Stat_NO2}
\end{table}

\section{Baseline Models}\label{App_Baseline_Models}

\noindent (a) Temporal baseline models:

\begin{itemize}
    \item \textbf{Autoregressive Integrated Moving Average} (ARIMA) is a popular statistical method for time series forecasting \citep{box2015time}. The ARIMA$(p,d,q)$ framework tracks the linear trajectory in a $d$-order (non-negative integer) differenced stationary time series by combining $p$ historical values of the target series and $q$ prior forecast errors. We utilize the \textit{forecast} package in R statistical software to implement the ARIMA model.
    
    \item \textbf{Long-short Term Memory} (LSTM) networks, a recurrent neural networks (RNN) architecture, is suitable for modeling long-term dependencies in time series forecasting \citep{hochreiter1997long}. This framework utilizes a gating mechanism with the input gate, forget gate and output gate to regulate the flow of information as short-term and long-term memory. 
    
    \item \textbf{Temporal Convolutional Network} (TCN) combines causal convolutions and dilated convolutions to model the long-term dependencies in a time series dataset \citep{chen2020probabilistic}. This architecture has a stable training mechanism due to skip connections in the residual blocks. 
    
    \item \textbf{DeepAR} is a variant of the RNN approach, capable of performing probabilistic forecasting \citep{salinas2020deepar}. This scalable architecture is suitable for handling complex seasonality in multiple time series.  
    
    \item \textbf{Transformers} is a state-of-the-art deep learning architecture that models complex patterns in time series data \citep{wu2020deep}. This framework utilizes the multi-head attention mechanism to capture the crucial information in a sequential learning problem. 
    
    \item \textbf{Neural Basis Expansion Analysis for Time Series} (NBeats) is a fully connected neural network architecture designed for time series forecasting \citep{oreshkin2019n}. This model comprises several blocks equipped with a basis expansion mechanism for transforming data into high-dimensional space and dense layers. The initial layers of the block are used for modeling and predicting past and future observations, while the subsequent layers are designed to remodel the errors and adjust the forecasts. 
\end{itemize}

To implement the above-mentioned deep learning models, we have used the \textit{darts} library in Python \citep{herzen2022darts}.

\noindent (b) Spatiotemporal baseline models:

\begin{itemize}
    \item \textbf{Space-time Autoregressive Moving Average} (STARMA) is a modification of the autoregressive moving average framework that incorporates the spatiotemporal auto-correlations \citep{pfeifer1980three}. This architecture includes both autoregressive and moving average terms that are lagged in space and time, making it useful for modeling linear trajectories in spatiotemporal systems. 
    
    \item \textbf{Generalized Space-time Autoregressive} (GSTAR) model is a robust spatiotemporal forecasting framework that allows the autoregressive parameters to vary across different locations \citep{cliff1975model, ruchjana2012least}. Unlike the STARMA model, the non-uniform weights of the GSTAR architecture make it more useful for modeling heterogeneous characteristics of sample locations. 
    
    \item \textbf{Fast Gaussian Process} (GpGp) method is a modification of Vecchia's Gaussian process approximation \citep{vecchia1988estimation}, designed for analyzing ordered sequences in time series observations \citep{guinness2018permutation}. This approach introduces a grouping mechanism for the ordered sequence, which significantly reduces the computational complexity of traditional Gaussian process forecasting methods. 

        \item \textbf{Spatiotemporal Graph Convolution Network} (STGCN) is a graph-based deep learning framework for performing spatiotemporal forecasting \citep{Yu_2018}. This framework comprises two spatiotemporal blocks, each containing a spatial graph convolution layer and two temporal gated convolution layers. The output from the spatiotemporal blocks is modeled with a fully connected dense layer to generate the required forecasts. Since this model combines multiple convolutional layers, it allows for faster training with fewer parameters.

    \item \textbf{Spatiotemporal Neural Network} (STNN) is a hybrid forecasting approach that combines the classical STARMA model with Support Vector Machine (SVM) and Artificial Neural Networks (ANN) to enhance forecast accuracy \citep{saha2020hybrid}. The STNN operates as an error-remodeling approach, where the original training data is first modeled using the linear STARMA model. 
    Residuals of the STARMA model are then modeled using the SVM architecture, and an ANN is subsequently applied to the predicted values of the residuals from the SVM to capture the nonlinearities better. The final STNN forecasts are obtained by aggregating the predictions from both the STARMA and ANN frameworks. 

    \item {\color{black}\textbf{Modified STGCN} serves as an intermediate design between the standard STGCN \citep{Yu_2018} and the proposed E-STGCN (developed in this study). The key difference between the modified version and the standard STGCN lies in the use of LSTM units for modeling temporal dependencies. In contrast, its primary difference from E-STGCN is based on its training mechanism, as discussed in the main text of the paper. Comparing the performance of E-STGCN with modified STGCN enables us to assess the impact of the EVT-based loss function on forecasting performance, particularly in the presence of extreme pollution events.}
        
    \item \textbf{Space-Time DeepKriging} (DeepKriging) model is a distribution-free spatiotemporal modeling architecture that is well-suited for handling non-Gaussian and non-stationary processes \citep{nag2023spatio}. The DeepKriging framework follows a two-step workflow, where radial basis functions and Gaussian kernels capture spatial and temporal trends, respectively. These spatiotemporal basis functions encode the coordinates, enabling effective spatiotemporal interpolation. In the second stage, convolutional LSTM networks are employed to generate forecasts based on the previously learned embeddings, allowing for more accurate spatiotemporal predictions.
\end{itemize}

To implement the STARMA, GSTAR, GpGp, and STNN models, we utilize the \textit{starma}, \textit{gstar}, \textit{GpGp}, and \textit{TDSTNN} packages in the R statistical software, respectively. We adopt the available implementation provided in the STGCN and DeepKriging model in \cite{Yu_2018} and \cite{nag2023spatio}, respectively. For the modified STGCN, we utilize a similar implementation as E-STGCN; however, the learning mechanism is performed solely based on MSE loss.

\section{MCB Plots}\label{App_MCB_Plots}

\begin{figure}[!ht]
    \centering
    \includegraphics[width = 1.0\textwidth]{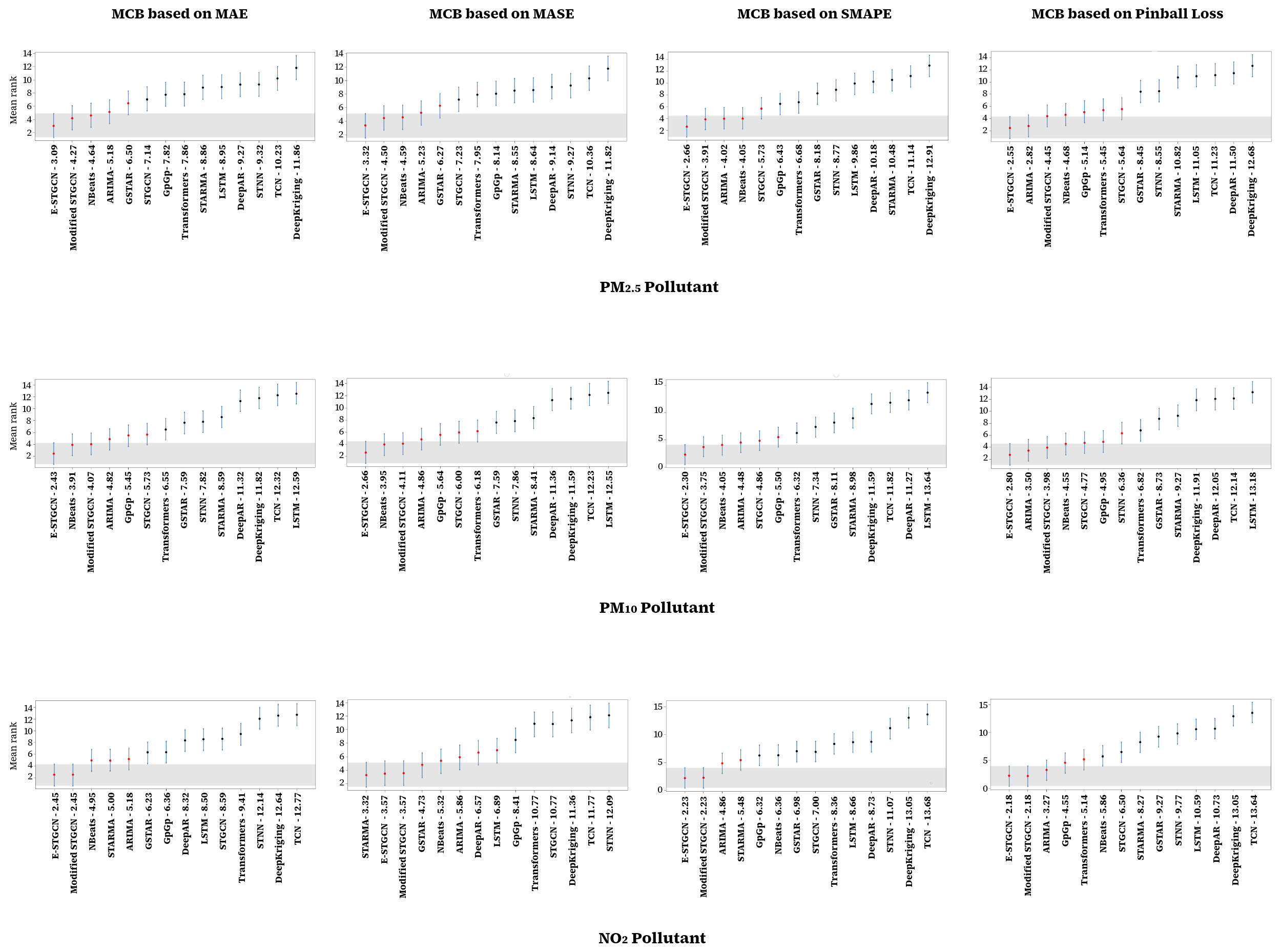}
    \caption{MCB plot for forecasting pollutant concentrations based on different performance metrics. In the figure, for example, `E-STGCN-3.09' means that the average rank of the proposed E-STGCN algorithm for $\PM_{2.5}$ forecasting, based on the MAE error metric, is 3.09; the same explanation applies to other algorithms, metrics, and pollutants. The shaded region depicts the reference value of the test.}
    \label{Fig_MCB_Pm25}
\end{figure}

The MCB test results for $\PM_{2.5}$, $\PM_{10}$, and $\NO_{2}$ pollutants are summarized in Fig. \ref{Fig_MCB_Pm25}. The results consider evaluation metrics MAE, MASE, SMAPE, and the quantile-based Pinball loss as discussed in Section \ref{Section_Performance_Metrics} of the main manuscript. For $\PM_{2.5}$, the MCB test results show that the E-STGCN framework achieves the lowest mean rank with values of 3.09 (MAE), 3.32 (MASE), 2.66 (SMAPE), and 2.55 (Pinball loss), followed by the modified STGCN, NBeats, ARIMA, and GSTAR models. The critical distance values of the remaining baseline models lie above the reference value (shaded region), indicating that their performance is significantly worse than the `best-fitted' E-STGCN model. For $\PM_{10}$, the E-STGCN framework consistently ranks as the `best' model across all performance indicators, followed by NBeats, modified STGCN, ARIMA, and GpGp. The performance of the other models is significantly inferior to the E-STGCN framework. In the case of forecasting $\NO_{2}$ levels, the E-STGCN and modified STGCN frameworks achieved similar rankings, emerging as the `best' models across all metrics except MASE, where STARMA performed best. Among the remaining models, NBeats, ARIMA, GSTAR, and GpGp performed significantly better than the other approaches. 

\end{document}